\documentclass[smallextended]{svjour3}

\usepackage{amssymb}
\usepackage{wrapfig}
\usepackage{rotating}
\usepackage{multirow}
\usepackage{array}
\usepackage{amssymb}
\usepackage{amsfonts}
\usepackage{graphicx}
\usepackage{epstopdf}
\usepackage{amsmath}
\usepackage{booktabs}
\usepackage{epigraph}
\usepackage{listings}
\usepackage{flushend}
\usepackage{url}
\usepackage{lscape}
\usepackage{xcolor}
\usepackage{comment}
\usepackage[utf8]{inputenc}
\usepackage{xspace}

\usepackage[pass]{geometry}

\newcounter{num_eq}
\usepackage[numbers]{natbib}
\usepackage{url}

\usepackage{color, colortbl}
\definecolor{Gray}{gray}{0.9}
\definecolor{White}{rgb}{255,255,255}

\newcommand{\MyPara}[1]{\vspace{.2em}\noindent\textit{\textbf{#1}}\hspace{.3em}}

\newcommand{\MyBox}[1]{\vspace{3mm}\noindent\framebox[\columnwidth][c]{\parbox[b]{0.95\columnwidth}{ #1 }}\vspace{3mm}}

\newboolean{showcomments}
\setboolean{showcomments}{true} 
\ifthenelse{\boolean{showcomments}}
 {
 
 }
 {\newcommand{\nbb}[2]{}
 
 }

\newcommand{\responsetoreviewer}[3][black]{#3}%

\newcommand{\responseinner}[1]{#1} 

\newcommand{\boldification}[1]{}

\begin{document}





\title{Tag that issue: Applying API-domain labels in issue tracking systems}


\titlerunning{Tag that issue: Applying API-domain labels in issue tracking systems}     

\author{
Fabio Santos . 
Joseph Vargovich . 
Bianca Trinkenreich . 
Italo Santos . 
Jacob Penney . 
Ricardo Britto . 
João Felipe Pimentel . 
Igor Wiese . 
Igor Steinmacher . 
Anita Sarma . 
Marco A. Gerosa
}

\authorrunning{Fabio Santos et al 

} 

\institute{Fabio Santos, Joseph Vargovich, Bianca Trinkenreich, Italo Santos, Jacob Penney, João Felipe Pimentel, Igor Steinmacher, Marco A. Gerosa  \at
                Northern Arizona Unversity \\
              \email{fabio\_santos@nau.edu, joseph\_vargovich@nau.edu, bianca\_trinkenreich@nau.edu, italo\_santos@nau.edu,  jacob\_penney@nau.edu, joao.pimentel@nau.edu, igor.steinmacher@nau.edu, marco.gerosa@nau.edu}           
              \and
         Igor Wiese,  \at
         Universidade Tecnológica Federal do Paraná \\
              \email{igor@utfpr.edu.br}
              \and
         Anita Sarma \at
         Oregon State University \\
              \email{anita.sarma@oregonstate.edu}
              \and
        Ricardo Britto \at
        Ericsson - Blekinge Institute of Technology \\
        \email{ricardo.britto@ericsson.com}
}

\date{Received: date / Accepted: date}


\maketitle
\begin{abstract}

Labeling issues with the skills required to complete them \responsetoreviewer{RevThreeMinorOne}{can help contributors to choose tasks} in Open Source Software projects. However, manually labeling issues is time-consuming and error-prone, and current automated approaches are mostly limited to classifying issues \responsetoreviewer{RevThreeMinorTwo}{as bugs/non-bugs}. We investigate the feasibility and relevance of automatically labeling issues with what we call ``API-domains,'' which are high-level categories of APIs. Therefore, we posit that the APIs used in the source code affected by an issue can be a proxy for the type of skills (e.g., DB, security, UI) needed to work on the issue. We ran a user study (n=74) to assess API-domain labels' relevancy to potential contributors, leveraged the issues' descriptions and the project history to build prediction models, and validated the predictions with contributors (n=20) of the projects. Our results show that (i) newcomers to the project consider API-domain labels useful in choosing tasks, (ii) labels can be predicted with a precision of 84\% and a recall of 78.6\% on average, (iii) the results of the predictions reached up to 71.3\% in precision and 52.5\% in recall when training with a project and testing in another (transfer learning), and (iv) project contributors consider most of the predictions helpful in identifying needed skills. These findings suggest our approach can be applied in practice to automatically label issues, assisting developers in finding tasks that better match their skills. 


\end{abstract}


\section{Introduction}\label{sec:intro}



Choosing a task to contribute to in Open Source Software (OSS) projects can be challenging~\cite{wang2011bug,steinmacher2015understanding,steinmacher2015systematic,10.1145/2675133.2675215,stanik2018simple}. Open tasks are publically reported in issue trackers, but since issues vary in complexity and required skills, contributors may find it difficult to select an appropriate task to undertake, especially when the contributors are new in the projects~\cite{zimmermann2010makes,Bettenburg:2007:QBR:1328279.1328284,vaz2019empirical,santoshits}. Adding labels to the issues (a.k.a. ``tasks,'' ``tickets,'' and ``bug reports'') is an effective way to help new contributors choose where to focus their efforts~\cite{steinmacher2018let}. \responsetoreviewer{RevOneCommentThree}{\responsetoreviewer[black]{RevTwoCommentTwo}{The labeling strategy supports a variety of contributors, including newcomers (new contributors), frequent contributors, and maintainers, as they have similar perceptions of the importance of this strategy \cite{santos2022how}.} Developers are newcomers each time they start a new project, no matter their previous experience.} Nevertheless, community managers and project maintainers find manually labeling issues challenging and time-consuming~\cite{9057411}. 


\responsetoreviewer{RevTwoCommentOneOuter}{We posit that the underlying APIs (the libraries required and imported into the source code) can be parsed to indicate skills required to work on an issue. \responsetoreviewer[black]{RevTwoCommentOne}{APIs are defined as ``a set of functions and procedures that enable the creation of applications that access the resources or data of an operating system, application or other services'' \cite{API-def}.} If the contributors know what types of APIs are used in the code to solve the issue, they could choose tasks that better match their skills or involve skills they want to learn. We leverage the idea that APIs encapsulate modules with specific purposes (e.g., cryptography, database access, logging) and abstract the details from the implementation. In this study, we focus on API-domain labels: high-level labels designating categories of APIs such as ``UI,'' ``Security,'' and ``Test,'' which may relate to skills needed to work on the issues.}



This paper extends our prior work~\cite{santos2021can}, in which we conducted a case study with a single project to investigate the feasibility of automatically labeling issues with API-domain labels. 
\responsetoreviewer{RevTwoCommentFive}{After running the first predictions with the case study, we observed that the number of dataset rows dropped significantly because of the lack of information about linked issues and PRs. With this in mind and to improve generalization, we looked for feasible ways to increase the datasets when a project is seriously affected by the dataset size after discarding issues not linked with a PR. In addition, it is sometimes impossible to access the source due to confidentiality in industry projects. Pursuing this reasoning, we sought ways to keep predicting the API-domain labels even when no training data is available by transferring the learning. Therefore, we believe the API-domain labels should be even more helpful if they can be applied in many open-source projects or industry projects despite their source code's dataset size or availability.} 
We extend the work by (1) expanding our study to five projects with diverse programming languages, vocabularies (natural languages), and issue track systems (ITS), (2) adding the BERT technique to our approach, (3) extending the qualitative analysis, (4) exploring the ability to transfer learning across projects, and (5) evaluating the API-domain labels with developers who solved the issues.

We answer the following research questions:

\MyBox{ 
\textbf{\emph{RQ.1:}} 
How relevant are the API-domain labels to new contributors?
\newline
\textbf{\emph{RQ.2:}} 
To what extent can we automatically attribute API-domain labels to issues?
\newline
\textbf{\emph{RQ.2.1:}} 
To what extent can we automatically attribute API-domain labels to issues using data from the project?
\newline
\textbf{\emph{RQ.2.2:}} 
To what extent can we automatically attribute API-domain labels to issues using data from \responsetoreviewer{RevThreeMinorThree}{other projects}?
\newline
\textbf{\emph{RQ.2.3:}} 
To what extent can we automatically attribute API-domain labels to issues using transfer learning?
\newline
\textbf{\emph{RQ.3:}} How well do the API-domain labels match the skills needed to solve an issue?
}

\responsetoreviewer{RevTwoCommentThree}{This paper studies the relevance of this labeling strategy to new contributors (RQ.1). We created models and evaluated their performance. Usually, machine learning approaches train and test data with the same project (RQ.2.1). However, when the existing data is not enough to create a prediction model with the expected performance, one may consider enlarging the dataset to include other projects (RQ.2.2). In addition, with the total absence of historical data for training in a target project, one can use a pre-trained dataset in the same domain (source project) to run predictions in the target project \cite{nam2013transfer} (RQ.2.3). Therefore, we also conducted transfer learning studies. Finally, the developers' opinions about the predictions were studied to determine whether the API-domain labels adequately indicate skills and help newcomers choose their tasks (RQ.3).}

Our contribution includes (1) how newcomers see the relevance of the API-domain labels; 
(2) a new semi-automated API classification process; (3) a mechanism to predict skills needed for projects coded in diverse programming languages (C, C\#, and Java), with issues in Portuguese and English; and (4) the validation of the API-domain labels with developers. 

\section{Related Work}
\label{sec:related}

Organizing issues involve some labeling efforts. Labeling is important for describing features and making it easier and faster to understand and search through software artifacts~\cite{santos2022how}. However, manually labeling software artifacts can be difficult and time-consuming. Thus, some approaches have been proposed for automatically labeling software projects \citep{Izadi2021TopicRF} and dependencies \citep{vargas2015automated}. While these approaches demonstrate the possibility of labeling software artifacts, they work at a higher level of classifying the whole project. In contrast, our approach classifies minor software artifacts (i.e., issues and APIs).

\responsetoreviewer{RevThreeMinorFour}{Approaches have also been proposed for labeling other software artifacts, such as questions from Stack Overflow} \citep{xia2013tag, lin2019pattern, 8643972}. \citet{xia2013tag} recommend tags for questions based on the similarity with previous questions. Those approaches are restricted to using only the existing tags and do not work with issue-tracking systems or APIs. \citet{8643972} and \citet{lin2019pattern} label opinions from users about APIs. Despite their focus on APIs, their goal is to support the developers' decisions to adopt a new API. In this work, we have the opposite goal. Given that a project already has APIs in different domains, our goal is to enable developers to find tasks that include APIs with which they are more familiar.

While many approaches are designed to label issues, most of them only try to distinguish bug reports from non-bug reports \citep{antoniol2008bug, pingclasai2013classifying, zhou2016combining, zhu2019bug, el2020automatic, perez2021bug}. Few approaches can classify according to other labels \citep{kallis2019ticket, izadi2022predicting, WANG2021107476}. Among them, \citet{izadi2022predicting} and \citet{WANG2021107476} use the text classification algorithm BERT \citep{Devlin2019BERTPO} for multiple labels, which we also use. Despite their ability to classify into distinct labels, such approaches only use pre-existing labels for classification. Instead of using predefined labels extracted from the existing issues or provided by default on GitHub, our approach define labels based on API domains. This kind of labeling helps to guide new contributors toward what to contribute \citep{Park.Jensen_2009,steinmacher2018let}, which can be a daunting task without guidance~\cite{steinmacher2015understanding}.

With a similar goal to support new contributors, social coding platforms like GitHub\footnote{\url{http://bit.ly/NewToOSS}} encourage projects to label issues\footnote{\responsetoreviewer{RevThreeCommentTenPartD}{In this study, the words ``tasks'' and ``issues'' are used interchangeably.}} that are easy for new contributors, which is done by several communities (e.g., LibreOffice,\footnote{\url{https://wiki.documentfoundation.org/Development/EasyHacks}} KDE,\footnote{\url{https://community.kde.org/KDE/Junior_Jobs}}, and Mozilla\footnote{\url{https://wiki.mozilla.org/Good_first_bug}}). However, community managers argue that labeling issues manually is difficult and time-consuming~\cite{9057411}. For that reason, \citet{huang2021characterizing} proposes an approach for labeling good first issues. While this approach indicates easy issues for new contributors, it is as limited in the outcome as the approaches that only classify issues as bugs. In contrast, by labeling issues with domains of the APIs, our approach can support new contributors of different skill levels. 

\section{Method Overview}

This section presents an overview of how we answered the research questions.

\begin{figure*}[!htb]
\centering
\includegraphics[width=1\textwidth] {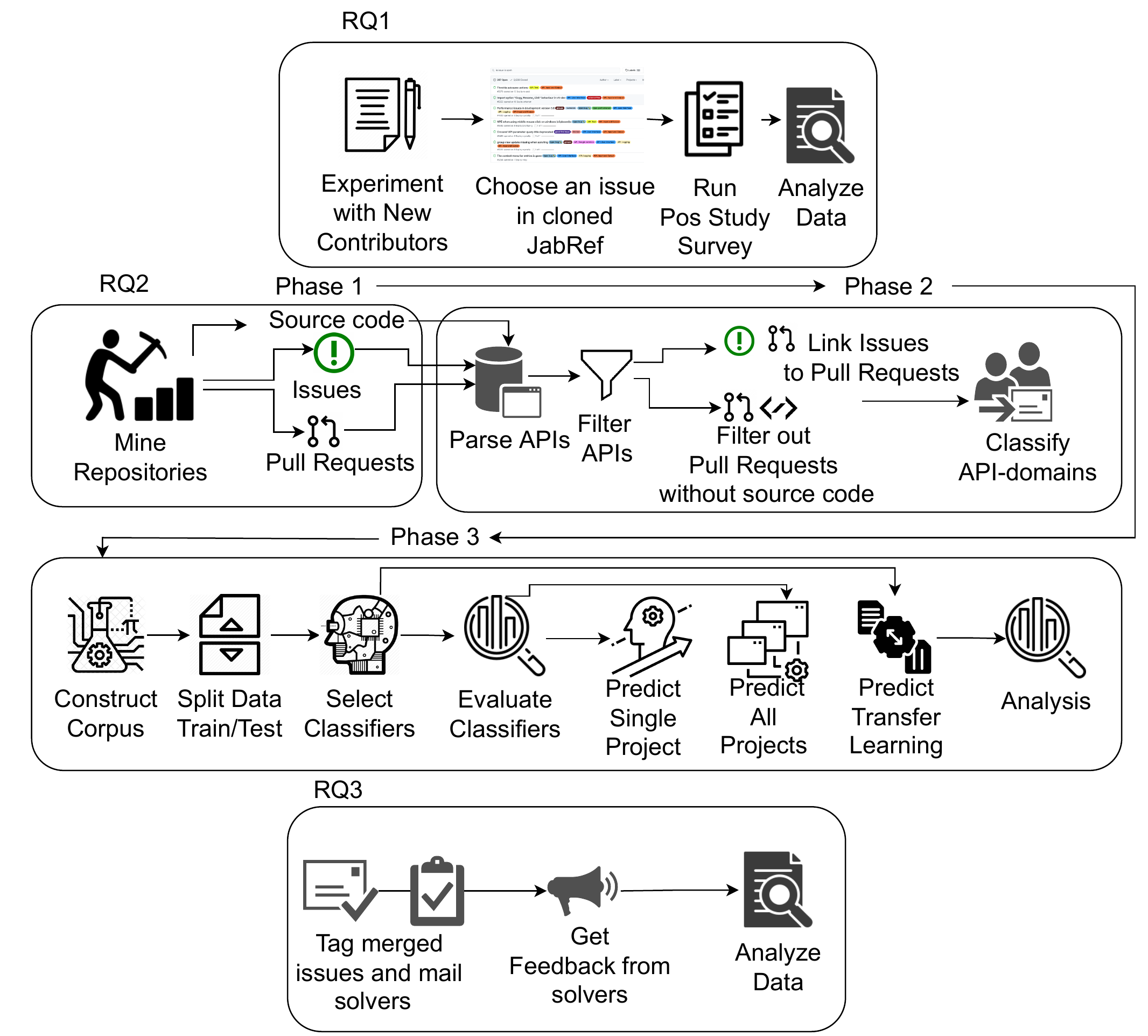}
\caption{\responsetoreviewer{RevThreeMinorSix}{Research method} overview}
\label{fig:researchdesignoverview}
\end{figure*}

\textbf{\emph{RQ.1:}} \textit{How relevant are the API-domain labels to new contributors?}
In this RQ (Section~\ref{sec:RQ1}), we evaluate the manually curated labels with potential new contributors. We divided the participants into two groups. After mimicking the project's issues pages for 22 issues, we added API-domain labels to the issues for the treatment group and kept the page as-is for the control group. We asked the participants to select issues to which to contribute and fill out a survey about their selection process (Figure~\ref{fig:researchdesignoverview} - RQ1). 

\textbf{\emph{RQ.2:}} 
\textit{To what extent can we automatically attribute API-domain labels to issues?}
In this RQ (Section~\ref{sec:RQ2}), we investigate the feasibility of predicting API-domain labels. We mined software repositories to collect issues, their associated pull requests, and the APIs used in the source code. Subsequently, we manually classified the APIs into API domains to build machine learning classifiers (Figure~\ref{fig:researchdesignoverview} - RQ2). To answer the sub-questions, we predicted the API-domain labels using each project dataset separately (RQ.2.1), a dataset with all projects merged (RQ.2.2), and different source and target datasets (RQ.2.3).



\textbf{\emph{RQ.3:}} \textit{How well do the API-domain labels match the skills needed to solve the issue?}
Finally, In this RQ (Section~\ref{sec:RQ3}), we asked contributors to provide feedback on the usefulness of the labels that we predicted in identifying skills needed to complete the issue (Figure~\ref{fig:researchdesignoverview} - RQ3). 

To foster reproducibility, we provide publicly available supplementary material\footnote{\url{https://doi.org/10.5281/zenodo.6869246}} containing the raw data, the Jupyter notebook scripts, and the anonymized survey data.

\section{Relevance of the Labels to \responsetoreviewer{RevTwoCommentSix}{New Contributors} (RQ1)} 
\label{sec:RQ1}

\subsection{Method} 
\label{sec:RQ1Method}

To explore the relevancy of the API-domain labels from an outsider's perspective, we conducted an experiment with 74 participants. We selected the JabRef project, hosted in GitHub, as the subject of the experiment. Two authors of this paper have already contributed to and have in-depth knowledge of the project. Having this knowledge helped us interpret the feedback and results. We created two versions of the JabRef issues page (with and without API-domain labels) and divided our participants into two groups (between-subjects design). We asked participants to choose and rank three issues to which they would like to contribute and answer a follow-up questionnaire about what information supported their decision. The artifacts used in this phase are part of the replication package.

\subsubsection{Participants}
\label{sec:Participants}


\responsetoreviewer{RevTwoCommentEight}{We used convenience sampling by recruiting participants from both industry and academia. We reached out to instructors and IT managers of our personal and professional networks and asked them to help in inviting participants. From industry, we recruited participants from one medium-sized IT startup hosted in Brazil and the IT department of a large and global company. We recruited students from multiple universities, including undergraduate and graduate computer science students from one university in the US and two others in Brazil as well as graduate data science students from a university in Brazil, since they are also potential contributors to the JabRef project.} Table~\ref{tab:demographics} presents the participants' demographics. We offered an Amazon Gift card (US\$ 25.00) to incentivize participation.

We categorized the participants' development tenure into novice and experienced coders, splitting our sample in half---below and above the average ``years as professional developer''. We also segmented the participants between industry practitioners and students. Participants are identified by a sequential number (column ``Participant'').

\begin{table}
\caption{Demographics Subgroups for the Experiment's Participants}

\centering
\label{tab:demographics}
\begin{tabular}{lrrllrr}
\cline{1-3} \cline{5-7}
\textbf{Population}     & \textbf{Quantity} & \textbf{Percentage} &  & \textbf{Tenure} & \textbf{Quantity} & \textbf{Percentage} \\ 
\cline{1-3} \cline{5-7} 
Industry & 41                     & 55.5 \%                  &  & Expert & 19                     & 25.7 \%                     \\

Student  & 33                     & 44.5 \%                  &  & Novice       & 55                     & 74.3 \%                     \\ 

\cline{1-3} \cline{5-7} 
\end{tabular}
\end{table}

The participants were randomly split into two groups: Control and Treatment. Out of the 120 participants that started the questionnaire, 74 (61.7\%) finished all the steps;  we only consider these participants in our analysis. We ended up with 33 and 41 participants in the Control and Treatment groups, respectively. 

\subsubsection{Experiment Planning}
\label{sec:ExperimentPlanning}

We selected 22 existing JabRef issues and built mock GitHub pages for Control and Treatment groups. \responsetoreviewer{RevTwoCommentSevenPartB}{The issues were selected from the most recent ones, trying to maintain similar distributions of the number of API-domain labels predicted per issue and the counts of predicted API-domain labels.} The control group mockup page had only the original labels from the JabRef issues, and the treatment group mockup page presented the original labels in addition to API-domain labels. These pages are available in the replication package. 
We used a preliminary version of our prediction model to generate the \responsetoreviewer{RevThreeMinorSeven}{API-domain labels \cite{santos2021can}}. 

\subsubsection{Questionnaire Data Collection}
\label{sec:SurveyDataCollection}

\begin{figure}
\centering
\includegraphics[width=1\textwidth]{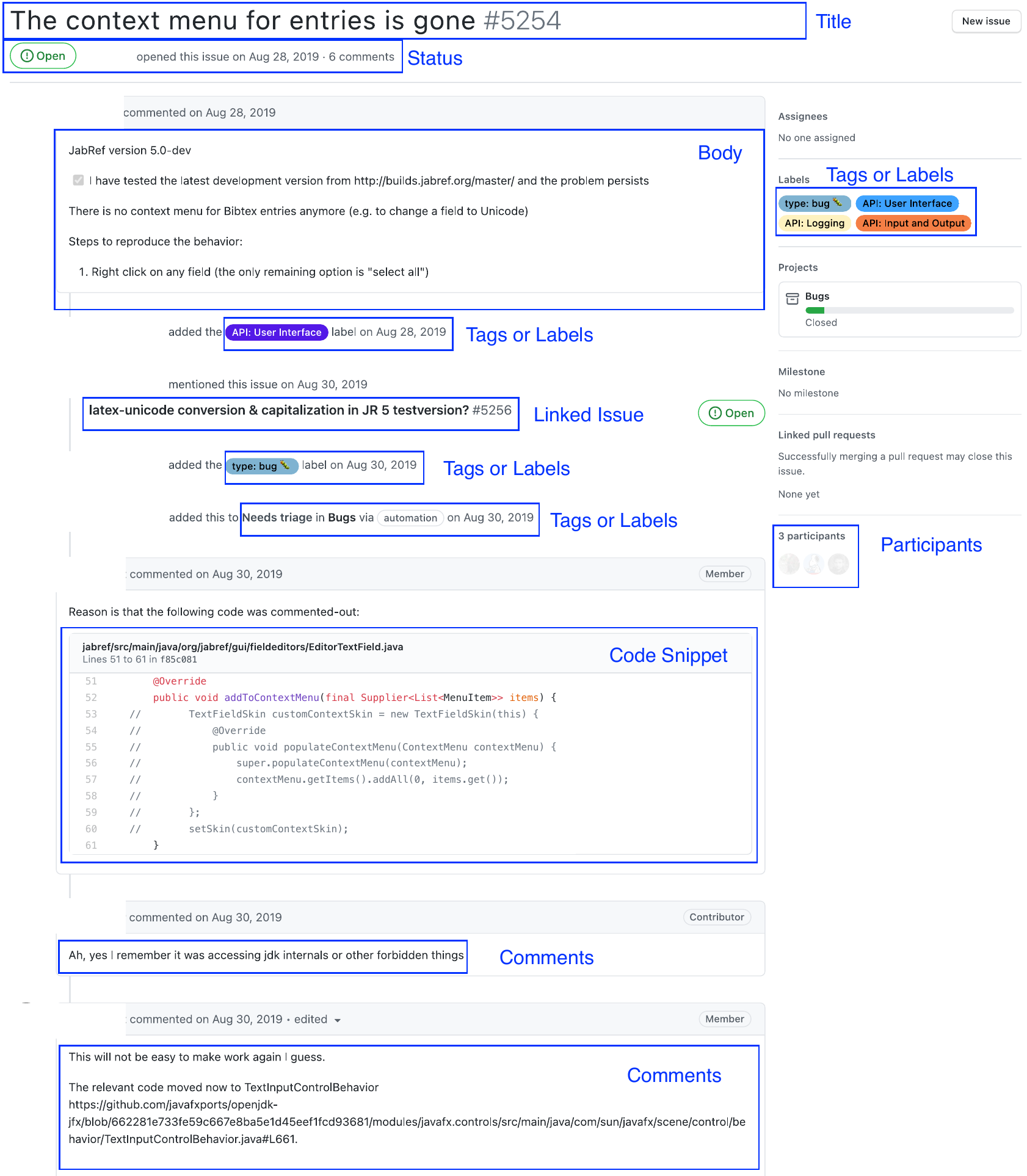}
\caption{Questionnaire question about the relevance of the page regions for task selection}
\label{fig:hotspotsurvey}
\end{figure}

The questionnaire included the following questions/instructions:

\begin{itemize}
 \item Select the three issues that you would like to work on.
 \item Select the information (region) from the issue page that helped you decide which issues to select (Fig:~\ref{fig:hotspotsurvey}).
 \item Why is the information you selected relevant? (open-ended question)
 \item Select the labels you considered relevant for choosing the three issues.

 \item What kind of label would you like to see in the issues? (open-ended question)
\end{itemize} 

The questionnaire also asked about participants' experience level, experience as an OSS contributor, and expertise level in the technologies used in JabRef.


Figure~\ref{fig:hotspotsurvey} shows an example of an issue details page and an issue entry on an issue list page. After selecting the issues to contribute, the participant was presented with this page to select what information region was relevant to their issue selection.

\subsubsection{Questionnaire Data Analysis}
\label{sec:SurveyDataAnalysis}

We split the analysis into two sets of questions.

\textbf{Regions and Labels Choices Analysis.}
\label{sec:RegionsDataAnalysis}
We first compared treatment and control groups' results to understand participants' perceptions about what information regions they considered important and the relevancy of the API-domain labels. We used violin plots to visually compare the distributions and measured the effect size using the Cliff's Delta test.

Then, we analyzed the data, aggregating participants according to their demographic information and resulting in the subgroups presented in Table~\ref{tab:demographics}. We calculated the odds ratio to check how likely it would be to receive similar responses from both groups. We used a 2x2 contingency table for each comparison---for instance, industry practitioners vs. students and experienced vs. novice coders. We used the following formula to calculate the odds ratio~\cite {szumilas2010explaining}: 

\begin{center}
\vspace{2mm}
$Odds Ratio (OR) = \frac{(a/c)}{(b/d)}$
\vspace{2mm}
\end{center}
An odds ratio $>$ 1 means that the first subgroup is more likely to report a type of label, while an odds ratio less than 1 means that the second group has greater chances (OR)~\cite {odds-ratio}.

\textbf{Open Questions Analysis.}
\label{sec:OpenDataAnalysis}
To understand the rationale behind the label choices, we qualitatively analyzed the answers to the open questions (``Why was the information you selected relevant?'' and ``What kind of label would you like to see in the issues?''). We selected representative quotes to illustrate the participants' perceptions of the labels' relevancy. 

We qualitatively analyzed the answers by inductively applying open coding in groups, where we identified the participant's reason for considering the provided information as relevant and what information the participant would like to be provided. We built post-formed codes as the analysis progressed and associated them with respective parts of the transcribed text to code the information relevance according to the participants' perspectives.

Researchers met weekly to discuss the coding. We discussed the codes and categorization until reaching a consensus about the meaning of and relationships among the codes. The outcome was a set of high-level categories as cataloged in our codebook\footnote{\url{https://doi.org/10.5281/zenodo.6869246}}.



\subsection{Results}
\label{sec:RQ1-results}


\boldification{Participants in the treatment group found labels more useful than the control group according to quantitative data}

\begin{figure}[!hbt]
\centering
\includegraphics[width=1\textwidth] {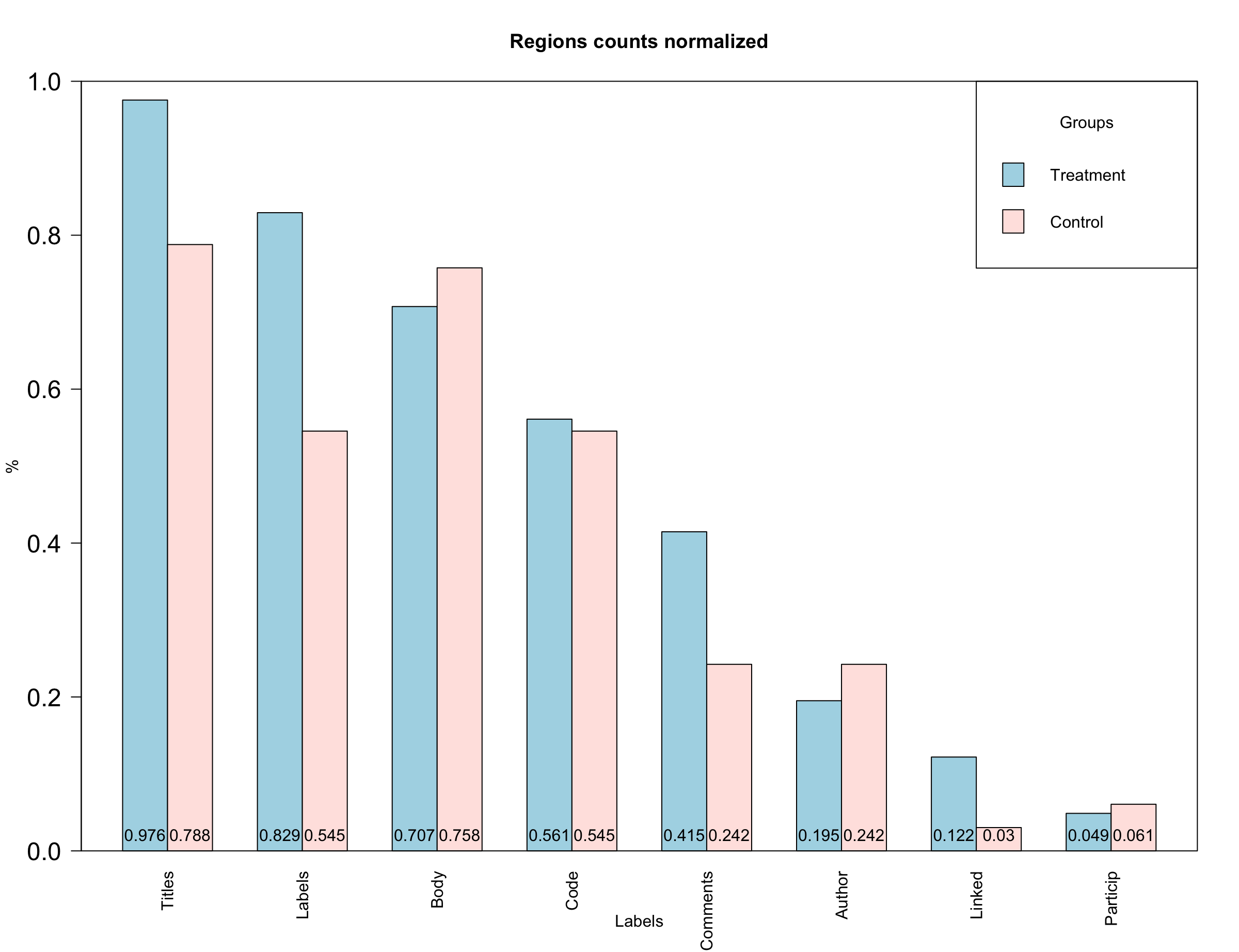}
\caption{The region counts (normalized) of the issue's information page selected as most relevant by participants from treatment and control groups.}
\label{fig:hotmapchoicesTC}
\end{figure}

\textbf{Information used when selecting a task.}
\label{sec:InformationTask}
Understanding the type of information that participants used in their decision to select an issue can help projects better organize such information on their issue pages. Figure~\ref{fig:hotmapchoicesTC} shows the different regions that participants found useful. In the control group, the top two regions of interest included the title of the issue (78.8\%) and the body (75.8\%), followed by the labels (54.5\%). 
This suggests that the labels used by the project were only marginally useful, and participants had to review the issue details. In contrast, in the treatment group, the top three regions of interest by priority were: title, label, and body 
97.6\%, 82.9\%, 70.7\%, 
respectively). This shows that participants in the Treatment group found the labels more useful than those in the control group: 82.9\% usage in the treatment group compared to 54.5\% in the control group. Comparing the body and the label regions in both groups, we found that participants from the treatment group selected 1.6x more label regions than the control group (p$<$0.05). 


\boldification{Qualitative data (quotes) reinforces this finding}

Our qualitative analysis reveals that the labels help in selecting issues. For instance, P2 
mentioned: \textit{``labels were useful to know the problem area and after reading the title of the issues, it was the first thing taken into consideration, even before opening to check the details''}. Participants found the labels to be useful in identifying the specific topic of the issue, as P4 
stated: \textit{``[labels are] hints about what areas have a connection with the problem occurring''}. 

\boldification{We also investigated which types of labels helped the most with decision-making: issue type, code component, or API-domain}

\textbf{Role of the labels}.
\label{roleLabels}
We also investigated which type of labels helped the participants in their decision-making. We divided the labels available to our participants into three groups based on the type of information. 

\begin{itemize}
\item Issue type (already existing in the project): This included information about the type of the task: bug, enhancement, feature, good first issue, and GSoC \responsetoreviewer{RevTwoCommentTen}{(Google Summer of Code)}.
\item Code component (already existing in the project): This included information about the specific code components of JabRef: entry, groups, external.files, main table, fetcher, entry.editor, preferences, import, keywords
\item API-domain (new labels): the labels generated by our classifier (IO, UI, network, security, etc.). These labels were available only to the treatment group. 
\end{itemize} 

\begin{table}[htb]
\centering

\caption{Label distributions among the control and treatment groups}
\label{tab:distributionLabels}
\begin{tabular}{lllll}
\hline
\textbf{Type of Label} & Control                 & \textbf{Percentage}                                          & \textbf{Treatment}               & \textbf{Percentage}                                           \\ \hline
Issue Type          & \multicolumn{1}{r}{145} & \multicolumn{1}{r}{56.4 \%}                      & \multicolumn{1}{r}{168} & \multicolumn{1}{r}{36.8 \%}                      \\
Components    & \multicolumn{1}{r}{112} & \multicolumn{1}{r}{43.6 \%} & \multicolumn{1}{r}{94}  & \multicolumn{1}{r}{20.6 \%} \\
API Domain          & \multicolumn{1}{r}{-}   & \multicolumn{1}{r}{-}                                             & \multicolumn{1}{r}{195} & \multicolumn{1}{r}{42.7 \%}                      \\ \hline
 
\end{tabular}
\end{table}

\boldification{API-domain labels are the preferred ones, as shown in table tab:distributionLabels}

Table~\ref{tab:distributionLabels} compares the labels that participants considered relevant (Section~\ref{sec:SurveyDataCollection}) across the treatment and control groups distributed across these label types. In the control group, the most selected labels (56.4\%) relate to the type of issue (e.g., Bug or Enhancement). In the treatment group, however, this number drops to 36.8\%, with API-domain labels as the majority (42.7\%), followed by code component labels (20.6\%). This difference in distributions alludes to the usefulness of the API-domain labels. 

\boldification{Comparing the frequency of choices of the three label types (fig:newlabels\_countsAC), API-domain labels are also the most frequently chosen}

\begin{figure}[!htb]
\centering
\includegraphics[width=.7\textwidth] {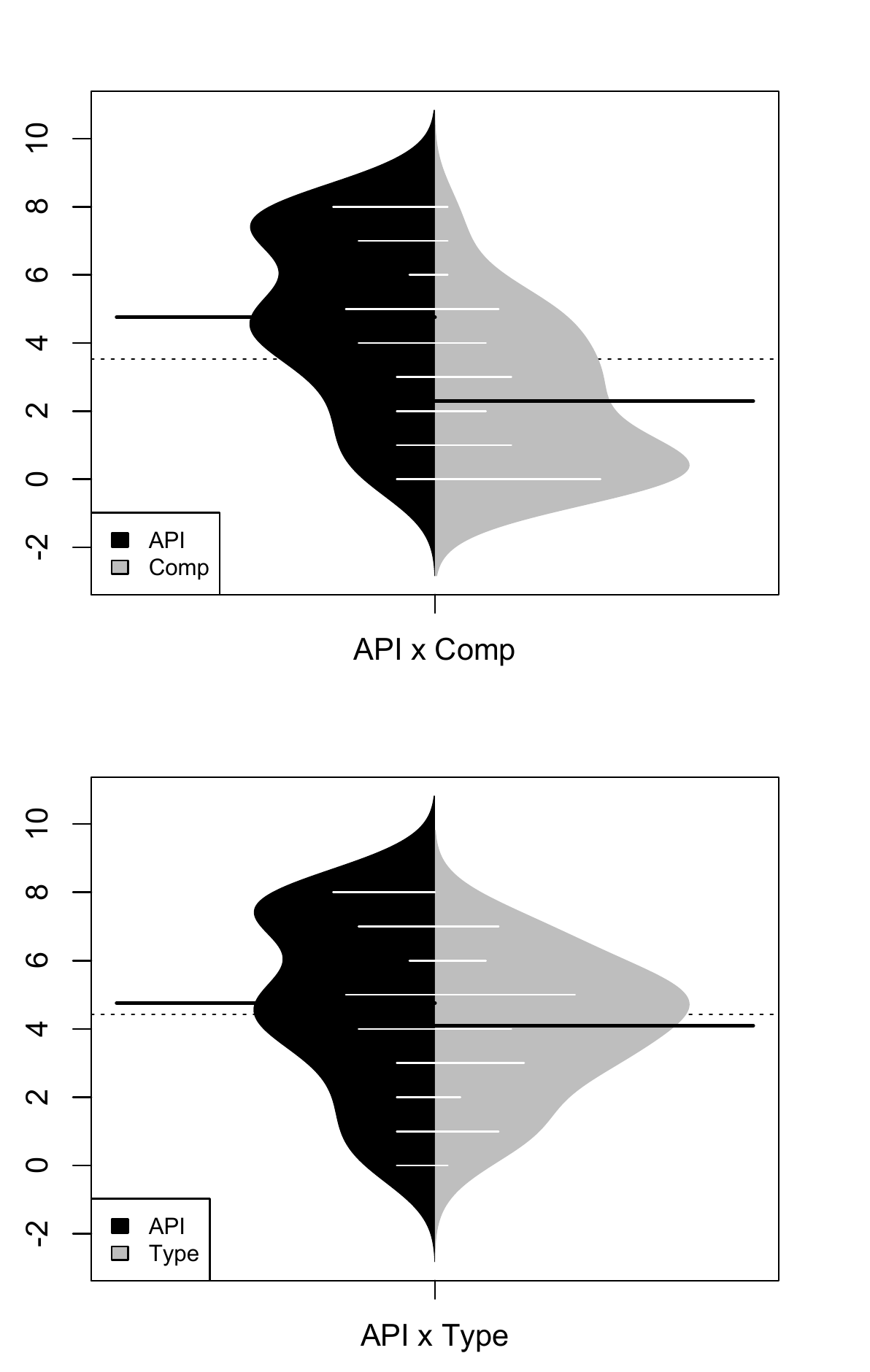}
\caption{\responsetoreviewer{RevThreeMinorEight}{The Y-Axis contains the density probability 
 and the median of} API-domain labels (API) x Component labels (Comp) x Type labels}
\label{fig:newlabels_countsAC}
\end{figure}

To better understand the usefulness of the API-domain labels as compared to the other types of labels, we further investigated the label choices among the treatment group participants. Figure~\ref{fig:newlabels_countsAC} presents two violin plots comparing (a) API-domain labels against code component labels and (b) API-domain labels against the type of issue. Wider sections of the violin plot represent a higher probability of observations taking a given value, the thinner sections correspond to a lower probability. The plots show that API-domain labels are more frequently chosen (median is 5 labels) as compared to code component labels (median is 2 labels), with a large effect size ($|$d$|$ = 0.52). However, the distribution of the issue type and API-domain labels are similar as confirmed by negligible effect size ($|$d$|$ = 0.1). These results indicate that while the type of issue (bug fix, enhancement, suitable for a newcomer) is important, understanding the technical (API) requirements of solving the task is equally important for developers deciding which task to select.


\boldification{We also analyzed the demography of the label choice. Industry participants preferred more API-domain labels than students. There was no statistical difference when comparing experienced coders vs. novice}

Finally, we analyzed whether the demographic subgroups held different perceptions about the API-domain labels (Table~\ref{tab:apiXcompXtype}). When comparing industry vs. students, we found participants from industry selected 1.9x (p-value=0.001) more API-domain labels than students when we controlled by component labels. We found the same odds when we controlled by issue type (p-value=0.0007). When we compared experienced vs. novice coders, we did not find statistical significance (p=0.11) when controlling by component labels. However, we found that experienced coders selected 1.7x more API-domain labels than novice coders (p-value=0.01) when we controlled by the type of the issue.

The odds ratio analysis suggests that API-domain labels are more likely to be perceived as relevant by practitioners and experienced developers than by students and novice coders.

\begin{table}
 \begin{center}
 \caption{Answers from different demographic subgroups regarding the API labels (API/Component/Issue Type)}
 \label{tab:apiXcompXtype}
 \begin{tabular}{c|c|r|r} 
  \hline
  \textbf{Subgroup} & \textbf{Comparison} & \textbf{API \%} & \textbf{Comp or Type \%} \\
  \hline

Industry & API/Comp & \textbf{56.0} & 44.0 \\
Students & API/Comp & 40.0 & \textbf{60.0} \\

Exp. Coders & API/Comp & \textbf{50.9} & 49.1 \\
Novice Coders & API/Comp & 41.5 & \textbf{58.5}  \\

Industry & API/issue Type & 45.5 & \textbf{55.5} \\
Students & API/issue Type & 30.6 & \textbf{69.4} \\

Exp. Coders & API/issue Type & 43.5 & \textbf{56.5} \\
Novice Coders & API/issue Type & 30.9 & \textbf{69.1}  \\

\hline
	
 \end{tabular}
 \end{center}
\end{table}

  







	



\boldification{We used the questionnaire to evaluate how subjects decided if a task was appropriate to them}

\textbf{The way contributors analyzed the issues.}\label{sec:5W2H} \responsetoreviewer{RevThreeCommentSix}{We used the questionnaire's open-ended question to evaluate how subjects used the information to decide whether the task was appropriate to them (Section \ref{sec:SurveyDataAnalysis})}.

\boldification{We found 22 categories of information based on the answers, which we organized using the 5W2H framework, that consists of 5-Wh and 2-How questions, as shown below}

\begin{figure*}[htb]
\centering
\includegraphics[width=1\textwidth]{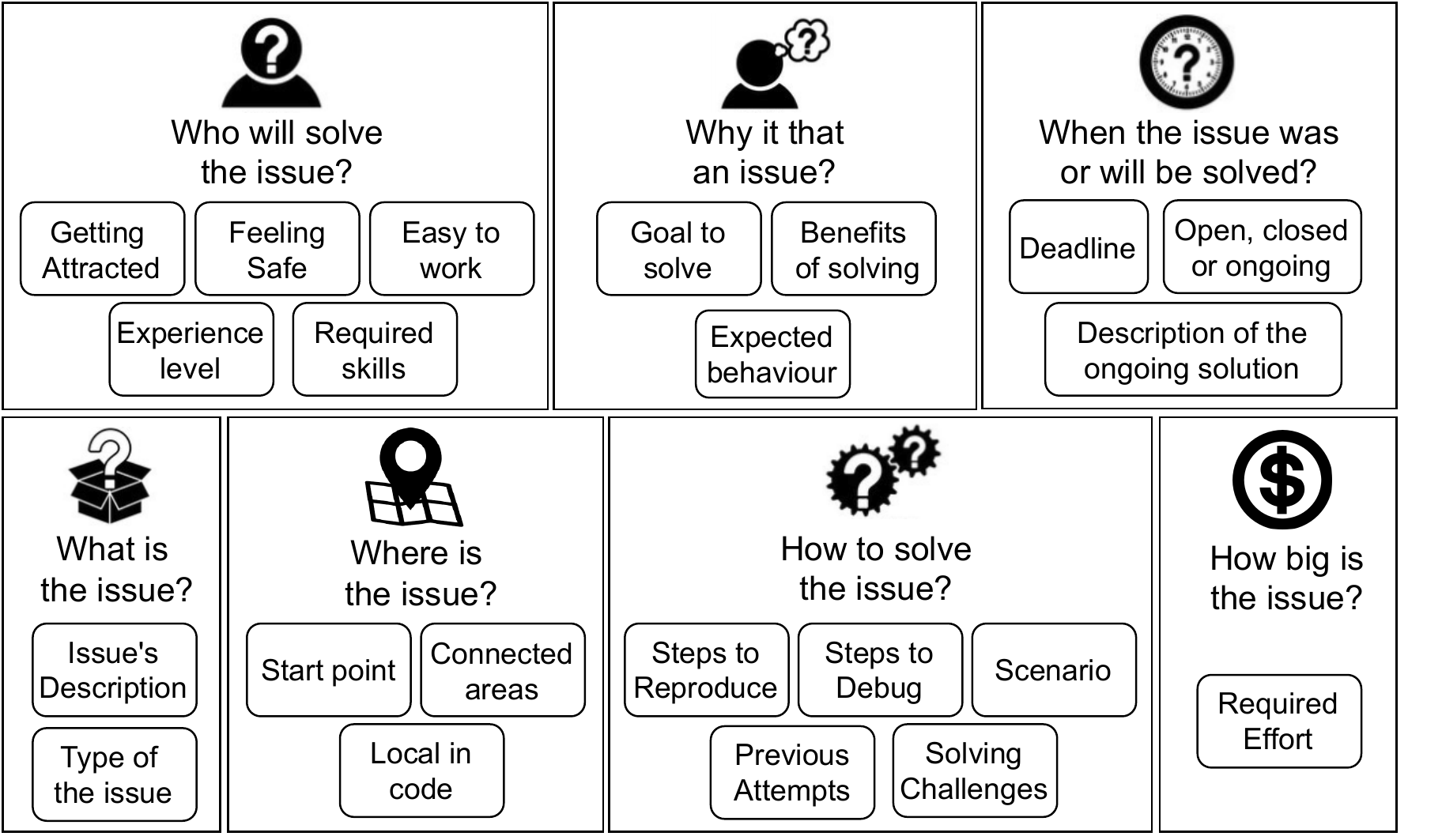}
\caption{The information reported by contributors as relevant to choosing a task. \responsetoreviewer{RevTwoCommentEleven}{We mapped the categories of our participants' definitions (rounded squares) to the 5W2H framework~\cite{klock20165w2h}, which organizes information for decision-making across seven questions.}}
\label{fig:results-rq1}
\end{figure*}

\responsetoreviewer{RevTwoCommentTwelvePartA}{Our qualitative analysis revealed a set of \responseinner{22} categories of information reported as relevant by contributors when they decide on a task to which to contribute. 
We organized the 22 categories of information based on an existing model from literature, the 5W2H framework, as we explain below and illustrate in Figure~\ref{fig:results-rq1}. The 5W2H framework (5-Wh and 2-How questions) is often used for clarifying a problem, issue, error, or nonconformity, or to facilitate implementing effective actions. The framework was initially applied to the automotive and other manufacturing industries~\cite{ohno1982toyota} and later to quality management~\cite{pacaiova2015analysis} and software engineering~\cite{klock20165w2h}.}

\responsetoreviewer{RevTwoCommentTwelvePartB}{\MyPara{\underline{Wh}o will solve the issue?} \responseinner{This category contains information about the forces influencing people to choose to work on an issue. Contributors mentioned what can influence one's decision to select the issue.} 
A newcomer can \textsc{become attracted} \responseinner{to select the issue} when \textit{``filtering labels to search issues that \responseinner{[they]} would like to contribute the most''} (P34) 
and reading the title (P18) 
to see if it \textit{``includes something that is not too wordy and if it uses words \responseinner{[they]} could easily understand''} (P21). 
When opening the issue, participants also reported the body and the comments were relevant to \textit{``gain interest on the issue''} (P4) 
, as a \textit{``detailed body and helpful comments from experienced people in the project is extremely helpful to make the newcomer \textsc{feeling safe} to try the issue''} (P6). 
The \responseinner{contributors' confidence to decide about an issue} can increase when they match \responseinner{their} \textsc{experience level} with the indication of difficulty to solve the issue (P8, P4) 
which could be shown in a label such as ``easy, medium, hard'' (P4) 
, ``good first issue'' (P6), 
or ``good challenging issue'' (P7). 
Besides \responseinner{their} experience, contributors can use the \textsc{required skills} to work on the issue to judge \textit{``if they have the \responseinner{[necessary]} skill to help''} (P31) 
or \textit{``whether or not [are] capable of finding a solution''} (P32). 
The required \responseinner{technical} skills mentioned by participants included the programming language of the code (P21, P27, P33), 
the architecture layer - front-end, back-end, interface (P27, P3, P2)
, APIs (P41, P42)
, database (P33)
, frameworks, and libraries (P20). 

\MyPara{\underline{Wh}y is that an issue?} \responseinner{is a category that justifies the issue as an issue. Participants mentioned the reasoning for an issue to exist could} raise interest in new contributors, such as knowing the \textsc{goal to solve} and \textit{``what is the purpose of the issue''} (P44)
, and the
\textsc{benefits of solving} or 
\textit{``why solving it will help users''} (P45). 
Additionally, the \textsc{expected behavior} of the software can help to clarify why the \responseinner{reported issue is an issue in comparison to the normal behavior of the software. Hence, the expected behavior represents a} \textit{``critical information to decide what is happening in the system and what is expected''} (P61). 
Indeed, one participant reported: \textit{``I would only contribute something that I know how it works''} (P22). 

\MyPara{\underline{Wh}en was the issue solved, or when will it be solved?}\responseinner{introduces time-related information and constraints regarding the issue. Participants reported they would like to know} the \textsc{deadline to solve} the issue 
\responsetoreviewer{RevThreeMinorNine}{or the \textit{``urgency''} (P13)}. 
Participants suggested that the priority appear in a label (P17) 
and be defined according to the impact that the issue has on businesses or users (P15). 
Another issue related to time is the \textit{``status to check the issue's state''} (P33), 
which can be \textsc{open, closed, or ongoing}, allowing contributors to use a filter in the issues' page. Since they ``don't look at closed issues much, [...] the open flag grabs [their] attention'' (P43). 
When a contributor is currently working on a solution, they should have their names assigned to the issue and include a comment with the \textsc{description of an ongoing solution} that should ``demonstrate the issue's status'' (P35). 


\MyPara{\underline{Wh}at is the issue?} \responseinner{relates to the description of the issue itself. Participants raised the importance of clear} \textsc{issues' description}, including both a summarized \textit{``idea of what the issue is about''} (P28) 
and a comprehensive explanation \textit{``to help understand what is the problem''} (P45) 
about. When an issue provides both levels of details, it \textit{``tells about the problem, first in a general term and later giving [them] details about it''} (P12). 
The issue's \textsc{type} in labels \textit{``demonstrate [...] how [the issue] is classified''} (P35). 
The participants suggested the issue should have \textit{``labels that inform precisely which type of issue is''} (P40): 
bug (P41)
, a new feature (P42)
, performance (P42)
, enhancement (P42)
, and security. One participant (P43) 
emphasized that \textit{```all issues should have a type so [they] can see if [their] skill set is useful''} (P43). 

\MyPara{\underline{Wh}ere is the issue?} \responseinner{references the localization of the issue in the code or project, guiding contributors to} a \textsc{start point} or \textit{``where to start looking at in the code/library to investigate the problem''} (P4). 
The \textsc{local in code} or the code block, method, or class which is causing the issue, 
and \textsc{connected areas}. This information would \textit{``give some hints about what areas have a connection with the problem occurring''} (P4) 
 and \textit{``code snippet to provide context for wherein the program this issue was happening''} (P18). 

\MyPara{\underline{H}ow to solve the issue?} \responseinner{brings practical directions to guide solving the issue. Awareness of \textit{``\textsc{previous attempts} to solve [an issue]''} (P30) 
helps contributors with} \textit{``valuable information about what has already been done and properly documented''} (P42). 
Contributors who are deciding about an issue can read \textit{``\textsc{[solving] challenges}''} (P35) 
\responseinner{to avoid wasting time on previous attempts and focus their effort on new paths to achieve} the solution. When working on the issue, having \textsc{steps to reproduce} the error (P45) 
on a controlled environment also help to solve the issue. \responseinner{Participants also mentioned they would like to see} \textit{``linked issues and comments to help understand the \textsc{scenario}''} (P33)
, and \textsc{steps to debug} to \textit{``decipher what the problems really is''} (P41). 

\MyPara{\underline{H}ow big is the issue?} is information that can provide visibility of the \textsc{required effort} for \textit{``[a contributor] to work on alone until [they] solve it''} (P7). 
If the issue does not have this information, the developer tries to \textit{``grasp what's the idea of the issue, to better measure how long it would take to solve it''} (P19). 

Finally, the question ``What region has this information?'' identifies the regions where the participants found the information in this study. Title appeared 7 times, body 8, comments 13, labels 5, status 2, code snippet 3, and linked issue 2.

\MyPara{5W2H outcomes:} the analysis confirmed the relevance of the title, body, comments, and labels and helped to create a taxonomy of what contributors analyze when deciding whether they want to contribute to an issue. The qualitative code we built for this open-ended question may be explored in future work to create ways to show the contributor such information using templates, labels, bots, or other UI objects.}\label{rev:RevTwoCommentTwelvePartC}

\textbf{Preferred types of labels.}
\label{sec:PreferredLabels}
Towards the evaluation of the labels contributors want to see in issues pages, 42 participants (out of 74) answered the open question Q2 (``What kind of label do you want to see in the issues?''). The \textsc{type}, \textsc{priority} to solve, and \textsc{programming language} \textit{``in which the code was written in''} (P21) 
were the three most mentioned, followed by \textsc{difficulty level}, \textsc{technology}, and \textsc{API}. 
Some participants suggested different semantics for the label \textsc{Type}: bug (P3, P41)
, improvement (P3)
, performance (P35, P42)
, new feature (P42)
, or security (P36). 
Other participants also suggested different semantics for \textsc{difficulty level}: \textit{``good first issue''} (P6)
, \textit{``good challenging issue''} (P7)
, or \textit{``easy, medium, hard''} (P4). 
The semantics for each label can be explored in future work. We present the 11 categories of suggested labels that we qualitatively coded from the participants' answers in Table~\ref{tab:desired_labels}.

\begin{table}[htb]
\centering
\caption{Labels desired by participants to select the issue}
\label{tab:desired_labels}

\begin{tabular}{
>{\columncolor[HTML]{FFFFFF}}l |
>{\columncolor[HTML]{FFFFFF}}l |
>{\columncolor[HTML]{FFFFFF}}l }
\hline
\textbf{Group}      & \textbf{Desired Labels}                                                                     & \textbf{Participants who Mentioned}                                                                                  \\ \hline
Management & Type                                                                               & \begin{tabular}[c]{@{}l@{}}P41, P35, P42, P3, P36, P37, P38, P39, P43, P40 
\end{tabular} \\ \hline
Management & Priority                                                                           & \begin{tabular}[c]{@{}l@{}} P12, P13, P19, P20, P14, P15, P16, P29, P17, P18
\end{tabular}  \\ \hline
Technical  & \begin{tabular}[c]{@{}l@{}}Programming \\ language\end{tabular}                    & \begin{tabular}[c]{@{}l@{}}P34, P33, P21, P22, P27, P23, P24, P22, P26
\end{tabular}         \\ \hline
Management & {\color[HTML]{333333} \begin{tabular}[c]{@{}l@{}}Difficulty \\ level\end{tabular}} & \begin{tabular}[c]{@{}l@{}}P4, P5, P6, P7, P9, P8
\end{tabular}                           \\ \hline
Technical  & Technology                                                                         & \begin{tabular}[c]{@{}l@{}}P30, P34, P33, P32, P20, P31
\end{tabular}                             \\ \hline
Technical  & API                                                                                & P41, P42, P3, P1, P20
\\ \hline
Technical  & \begin{tabular}[c]{@{}l@{}}Architecture \\ layer\end{tabular}                      & P2, P3, P27, P18
\\ \hline
Management & Status                                                                             & P28, P9, P29
\\ \hline
Management & GitHub info                                                                        & P10, P19
\\ \hline
Technical  & Database                                                                           & P33
\\ \hline
Technical  & Framework                                                                          & P20 
\\ \hline
\end{tabular}
\end{table}

\boldification{Labels about github info, database, and framework are less relevant}

Participants prefer to see labels on priority, type, programming language, complexity, technology, and APIs more than architecture, status, GitHub info, database, and framework. \responsetoreviewer{RevTwoCommentThirteen}{GitHub info is general information about the project repository (e.g., ``ranking about the most commented'' (P10P0) and ``branches'' (P22I0))}.

\MyBox{\textbf{\emph{RQ.1 Summary.}} Our findings suggest that labels are relevant for selecting an issue to work on. API-domain labels increased the perception of the labels' relevancy. API-domain labels are especially relevant for industry and experienced coders. API is one of the issue labels users want to see. 5W2H analysis has confirmed the relevance of labels and can guide contributors on how to write an issue.
} 


\section{Label Predictions (RQ2)}
\label{sec:RQ2}

Even with the relevance of the API-domain labels, we investigated how to predict them automatically.

\subsection{Method}
\label{sec:RQ2Method} 

To predict the API-domain labels, we employed a multi-label classification approach. This approach is divided into three phases: phase 1 - mining the repositories; phase 2 - parsing the source code and semi-automatically categorizing the APIs with experts; and phase 3 - building the corpus and running the classifiers (Figure~\ref{fig:researchdesignoverview}). Additionally, we explored the influence of issue elements (i.e., title, body, and comments) and machine learning setup (i.e., n-grams and different algorithms) on the predictions. 

\begin{table*}[t]
\centering
\caption{Project Details. R - Number of Releases. C - Number of Contributors}
 \label{tab:projectDetails}

\responsetoreviewer{RevTwoCommentFourteen}{
\begin{tabular}{l|r|r|r|r|r|r|l|c}
\hline
\textbf{Project} & \textbf{\begin{tabular}[c]{@{}r@{}}R\end{tabular}} & \textbf{\begin{tabular}[c]{@{}r@{}}C\end{tabular}} & \textbf{\begin{tabular}[c]{@{}r@{}}Stars\end{tabular}} & \textbf{\begin{tabular}[c]{@{}r@{}}Forks\end{tabular}} &\textbf{\begin{tabular}[c]{@{}r@{}}Closed\\ Pulls\end{tabular}} &\textbf{\begin{tabular}[c]{@{}r@{}} Issues\end{tabular}} & \textbf{Domain}                                                     & \textbf{OSS} \\ \hline
JabRef           & 42                                                            & 337                                                               & \begin{tabular}[c]{@{}r@{}}15.7K\end{tabular}     & \begin{tabular}[c]{@{}r@{}}1.8K\end{tabular}        & \begin{tabular}[c]{@{}r@{}}2.7K\end{tabular}      & \begin{tabular}[c]{@{}r@{}}4.1K\end{tabular}                  & \begin{tabular}[c]{@{}l@{}}Articles \\manager  \end{tabular}                                                  & Y            \\
Audacity         & 25                                                            & 154                                                               & \begin{tabular}[c]{@{}r@{}}6.9K\end{tabular}     & \begin{tabular}[c]{@{}r@{}}17.1K\end{tabular}       & \begin{tabular}[c]{@{}r@{}}1.1K\end{tabular}   & \begin{tabular}[c]{@{}r@{}}0.6K\end{tabular}                     & \begin{tabular}[c]{@{}l@{}}Audio \\editor   \end{tabular}                                                     & Y            \\
PowerToys        & 50                                                            & 262                                                               & \begin{tabular}[c]{@{}r@{}}65.5K\end{tabular}   & \begin{tabular}[c]{@{}r@{}}3.7K\end{tabular}          & \begin{tabular}[c]{@{}r@{}}3.3K\end{tabular}    & \begin{tabular}[c]{@{}r@{}}9.9K\end{tabular}                      & \begin{tabular}[c]{@{}l@{}}Utilities \\for Windows   \end{tabular}                                            & Y            \\
RTTS             & 121                                                           & 40                                                                & \begin{tabular}[c]{@{}r@{}}N.A.\end{tabular}  & \begin{tabular}[c]{@{}r@{}}N.A.\end{tabular}              & \begin{tabular}[c]{@{}r@{}}N.A.\end{tabular}   & \begin{tabular}[c]{@{}r@{}}N.A.\end{tabular}                      & \begin{tabular}[c]{@{}l@{}}Telecommu-\\nication\\ product\end{tabular} & N            \\
Cronos           & 123                                                           & --                                                                & \begin{tabular}[c]{@{}r@{}}N.A.\end{tabular}  & \begin{tabular}[c]{@{}r@{}}N.A.\end{tabular}             & \begin{tabular}[c]{@{}r@{}}N.A.\end{tabular}     & \begin{tabular}[c]{@{}r@{}}N.A.\end{tabular}                    & \begin{tabular}[c]{@{}l@{}}Time \\Tracker    \end{tabular}                                                    & N            \\ \hline
\end{tabular}

}
\end{table*}

\responsetoreviewer{RevThreeCommentSevenPartA}{In our preliminary work~\cite{santos2021can}, we conducted an exploratory experiment on a single project (JabRef). In the current study, we include four new projects. We selected projects to increase the diversity of domains, programming languages, and human languages (vocabularies). We sought a mix of popular open-source (OSS) and closed-source currently active projects with a large number of issues and pull requests. As we aimed to run surveys within the project communities, we contacted maintainers/managers of candidate projects in advance to explain our goals and seek support in reaching contributors for the user studies. Table~\ref{tab:projectDetails} presents the selected projects and their characteristics.} 

\responsetoreviewer{RevThreeCommentSevenPartB}{The study can be divided into two branches of prediction: TF-IDF and BERT. The TF-IDF predictions followed the previous study \cite{santos2021can}, employing five algorithms (Random Forest Decision Tree Logistic Regression, MLP Classifier, and MLkNN) but were extended to more projects, ITSs, programming languages, and vocabularies (natural languages). The BERT predictions operate the same extensions but are restricted to English vocabulary. Unlike the TF-IDF, BERT determines the meanings of words in a corpus based on their context within a sentence. 
We compared BERT to the previous TF-IDF classification pipeline within the context of the issue labeling problem.}



\subsubsection{Phase 1 - Mining Software Repositories}
\label{sec:DataCollection}


We started by gathering data from the repositories to train a machine learning model to predict the API labels. To achieve this goal, we mined closed issues and merged pull requests. 
Table~\ref{tab:projectsmined3} summarizes the projects' characteristics and demographics. 
\responsetoreviewer{RevTwoCommentSixteen}{We collected a total of 22,231 issues and 4,674 pull requests (PR) from all projects, considering all project data until November 2021}. For the OSS projects, we used the GitHub REST API v3 to collect data such as title, body, comments, and closure date. We also collected the name of the files changed in the PR and the commit message associated with each commit. The industry projects used Gerrit \responsetoreviewer{RevThreeMinorTen}{(RTTS - Real-Time Telecom Software) and Jira + MTT - Minds At Work Time Tracker (Cronos). From RTTS, we extracted two CSVs files: one containing the ``issues'' (troubles in RTTS) and the second containing the commits. The Cronos project uses a combination of Jira to track the open issues and the software MTT, an in-house solution, to manage the revisions and allocation time. We extracted a CSV file from Jira and a TXT from MTT.}


\begin{table}[ht]
\centering
\scriptsize
\caption{Projects Mined and Issue Tracker Systems}
 \label{tab:projectsmined3}

\begin{tabular}{l|lll|c|c|c|c}
\hline
\textbf{Project} & \multicolumn{1}{l|}{\textbf{\begin{tabular}[c]{@{}l@{}}Prog\\ Lang\end{tabular}}} & \multicolumn{1}{l|}{\textbf{\begin{tabular}[c]{@{}l@{}}Issue\\ Tracker/\\ Vocab-\\ ulary\end{tabular}}} & \textbf{\begin{tabular}[c]{@{}l@{}}Extraction\\ Method\end{tabular}} & \textbf{\begin{tabular}[c]{@{}c@{}}Issues/\\ PR\end{tabular}} & \textbf{\begin{tabular}[c]{@{}c@{}}Linked\\ Issues\\ \& PR\end{tabular}} & \textbf{\begin{tabular}[c]{@{}c@{}}Source\\ Code\\ Files\end{tabular}} & \textbf{\begin{tabular}[c]{@{}c@{}}Distinct\\ APIs\end{tabular}} \\ \hline
JabRef           & \multicolumn{1}{l|}{Java}                                                         & \multicolumn{1}{l|}{\begin{tabular}[c]{@{}l@{}}GitHub\\ EN\end{tabular}}                                & \begin{tabular}[c]{@{}l@{}}GitHub \\ API V3\end{tabular}             & \begin{tabular}[c]{@{}c@{}}4,471\\ 1,966\end{tabular}         & 1,914                                                                     & 1,690                                                                  & 1,944                                                            \\
Audacity         & \multicolumn{1}{l|}{C++}                                                          & \multicolumn{1}{l|}{\begin{tabular}[c]{@{}l@{}}GitHub\\ EN\end{tabular}}                                & \begin{tabular}[c]{@{}l@{}}GitHub \\ API V3\end{tabular}             & \begin{tabular}[c]{@{}c@{}}1,440\\ 310\end{tabular}           & 341                                                                      & 624                                                                    & 1,478                                                            \\
PowerToys        & \multicolumn{1}{l|}{C\#}                                                          & \multicolumn{1}{l|}{\begin{tabular}[c]{@{}l@{}}GitHub\\ EN\end{tabular}}                                & \begin{tabular}[c]{@{}l@{}}GitHub \\ API V3\end{tabular}             & \begin{tabular}[c]{@{}c@{}}12,571\\ 853\end{tabular}          & 1,011                                                                     & 794                                                                    & 264                                                              \\
RTTS             & \multicolumn{1}{l|}{Java}                                                         & \multicolumn{1}{l|}{\begin{tabular}[c]{@{}l@{}}Gerrit\\ EN\end{tabular}}                                & \begin{tabular}[c]{@{}l@{}}Export \\ CSV\end{tabular}                & \begin{tabular}[c]{@{}c@{}}2,836\\ 470\end{tabular}           & 470                                                                      & 9,779                                                                  & 8,645                                                            \\
Cronos           & \multicolumn{1}{l|}{Java}                                                         & \multicolumn{1}{l|}{\begin{tabular}[c]{@{}l@{}}Jira/MTT\\ BR\end{tabular}}                              & \begin{tabular}[c]{@{}l@{}}Export \\ CSV/TXT\end{tabular}            & \begin{tabular}[c]{@{}c@{}}913\\ 1075\end{tabular}            & 206                                                                      & 220                                                                    & 441                                                              \\ \hline
\textbf{Total}            & \multicolumn{3}{l|}{}                                                                                                                                                                                                                                              & \begin{tabular}[c]{@{}c@{}}22,231\\ 4,674\end{tabular}        & 3,942                                                                     & 13,107                                                                 & 12,772                                                           \\ \hline
\end{tabular}
\end{table}





Next, to train the model, we kept only the data from issues linked with merged and closed pull requests, since we needed to map issue data to source code APIs. To find the links between pull requests and issues in open source projects, we searched for the symbol \verb|#|issue\verb|_|number in the pull request title and body and checked the URL associated with each link. We also filtered out issues linked to pull requests without at least one source code file (e.g., those associated only with documentation files) since they do not provide the model with content related to any API. Similarly, we linked projects hosted by Gerrit and Jira/MTT, 
using the trouble ID and key fields (Gerrit), and for the project managed with Jira/MTT, we linked using the change ID and revision fields. The TXT file from MTT needed to be parsed to look for the revision field. We discarded entries without source code or linked data. In total, 734 entries were discarded. 




\subsubsection{Phase 2 - API classification} 
\label{sec:APIclassification} 

Phase 2 encompasses API extraction and expert classification.

\textbf{API extraction.}
\label{sec:APIextraction}
To identify the APIs used in the source code affected by each pull request, we built a parser to process all source files from the projects. In total, 12,772 library declaration statements from 13,107 source files were mapped to 185,159 possible relationships between files and APIs. The parser looked for specific commands, i.e., import (Java), using (C\#), and include (C++). The parser identified all classes, including the complete namespace from each import/using/include statement. We considered only the most frequent language per project. 

Then, we filtered out APIs not found in the latest version of the source code (JabRef 5.3, 
Audacity 3.1.0, and PowerToys 0.49.1; RTTS and Cronos are industry projects, and we used the last provided version) to avoid recommending APIs in source code that were no longer used in the project. \responsetoreviewer{RevTwoCommentTwentyTwo}{The filtering process is automatic. When processing a closed pull request, the files attached have their filenames compared with those stored in a database by the parser. When the file name is not found in the database, the pull request is discarded from the training set.}

Our final dataset comprises 22,231 issues, 4,674 pull requests, 13,107 files, and 12,772 distinct APIs (Table~\ref{tab:projectsmined3}). 


\begin{table}[htbp]

\centering
\caption{Labels Definition}
\label{tab:labelsdefinition}
\begin{tabular}{l|l}
\hline
\multicolumn{1}{c|}{\textbf{Label Generated}} & \multicolumn{1}{c}{\textbf{Definition}} \\ \hline \hline
Application (App) & \begin{tabular}[c]{@{}l@{}}Third-party apps or plugins for specific use attached to \\ the System\end{tabular} \\ \hline
\begin{tabular}[c]{@{}l@{}}Application Performance \\ Manager (APM)\end{tabular} & Monitors performance or benchmark \\ \hline
Big Data & \begin{tabular}[c]{@{}l@{}}APIs that deal with storing large amount of data, with\\  variety of formats\end{tabular} \\ \hline
Cloud & APIs for software and services that run on the Internet \\ \hline
Computer Graphics (CG) & Manipulating visual content \\ \hline
Data Structure & Data structures patterns (e.g., collections, lists, trees) \\ \hline
Databases (DB) & Databases or metadata \\ \hline
\begin{tabular}[c]{@{}l@{}}Software Development and \\ IT Operations (DevOps)\end{tabular} & \begin{tabular}[c]{@{}l@{}}Libraries for version control, continuous integration and \\  continuous delivery\end{tabular} \\ \hline
Error Handling & Response and recovery procedures from error conditions \\ \hline
Event Handling & Answers to events like listeners \\ \hline
\begin{tabular}[c]{@{}l@{}}Geographic Information \\ System (GIS)\end{tabular} & Geographically referenced information \\ \hline
Input-Output (IO) & Read, write data \\ \hline
Interpreter & Compiler or interpreter features \\ \hline
Internationalization (i18n) & \begin{tabular}[c]{@{}l@{}}Integrate and infuse international, intercultural, and \\  global dimensions\end{tabular} \\ \hline
Logic & \begin{tabular}[c]{@{}l@{}}Frameworks, Patterns like Commands, Controls or\\  architecture-oriented classes\end{tabular} \\ \hline
Language (Lang) & Internal language features and conversions \\ \hline
Logging & Log registry for the app \\ \hline
Machine Learning (ML) & ML support like build a model based on training data \\ \hline
Microservices/Services & \begin{tabular}[c]{@{}l@{}}Independently deployable smaller services. Interface \\ between two different applications so that they can \\ communicate with each other\end{tabular} \\ \hline
Multimedia & Representation of information with text, audio, video \\ \hline
Multi-Thread (Thread) & Support for concurrent execution \\ \hline
\begin{tabular}[c]{@{}l@{}}Natural Language \\ Processing (NLP)\end{tabular} & Process and analyze natural language data. \\ \hline
Network & Web protocols, sockets, RMI APIs \\ \hline
Operating System (OS) & APIs to access and manage a computer's resources \\ \hline
Parser & \begin{tabular}[c]{@{}l@{}}Breaks down data into recognized pieces for further \\  analysis.\end{tabular} \\ \hline
Search & API for web searching \\ \hline
Security & Crypto and secure protocols \\ \hline
Setup & Internal app configurations \\ \hline
User Interface (UI) & Defines forms, screens, visual controls \\ \hline
Utility (Util) & Third-party libraries for general use \\ \hline
Test & Test automation \\ \hline \hline
\end{tabular}

\end{table}

\textbf{Expert Classification.}
\label{sec:ExpertClassification}
\responsetoreviewer{RevTwoCommentSevenPartA}{Three software engineering experts (senior developers), including one of the authors of this article, proposed the labels based on their experience in software development, considering possible categories generic enough to suit a wide range of APIs present in software projects. For example, the proposed domains contain UI, IO, Cloud, Error handling, etc. After four rounds of discussions, the experts reached a consensus, and 31 API domains were defined (Table~\ref{tab:labelsdefinition})}. 

\begin{figure}[!hbt]
\centering
\includegraphics[width=1\textwidth] {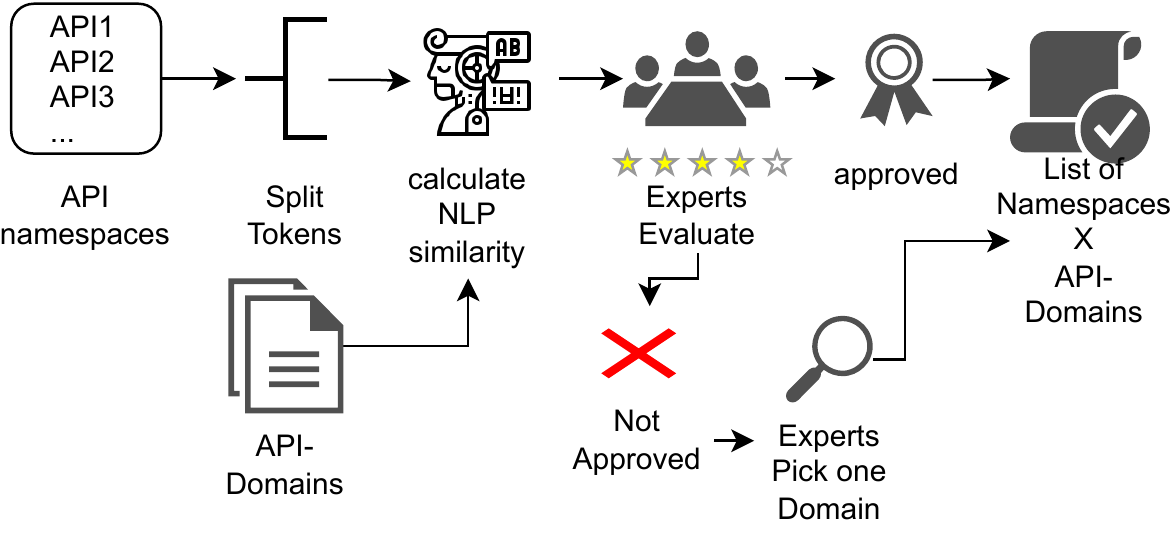}
\caption{Process for evaluating APIs by experts}
\label{fig:NLPExpert}
\end{figure}

After defining the 31 API domains, we started to classify the APIs semi-automatically (Figure~\ref{fig:NLPExpert}). The intuition behind the API classification method is that libraries' namespaces often reveal architectural information and, consequently, their categories or API domains~\cite{ducasse2009software, savidissoftware}. To identify the possible API domains for each API, we split all the API namespaces into tokens. For instance the API ``com.oracle.xml.util.XMLUtil'' was split in ``com'', ``oracle'', ``xml'', ``util'', and ``XMLUtil''. Next, we eliminated the business domain name extensions (e.g., ``org'', ``com''), country code top-level domain (``au'', ``uk'', etc.), and the project and company names (``microsoft'', ``google'', ``facebook'', etc.). In the example, we kept the first token ``xml'', second token ``util'', and full namespace ``com.oracle.xml.util.XMLUtil.'' 

For each token, we identified how similar it is to the 31 proposed API domains using an NLP similarity function. The intention is to suggest to the experts potential fits for the APIs. We used the NLP Python package spacy \responsetoreviewer{RevTwoCommentSeventeen}{\cite{spacy.io}}. Spacy is a multi-use NLP package and can retrieve the semantic similarity of words using word2vec. We set up the spacy package with the largest trained model available (large full vector package, en\_core\_web\_lg, which includes 685k unique vectors).

\responsetoreviewer{RevTwoCommentEighteen}{To assist the expert evaluation and reduce the search scope, we aggregated the tokens found in namespaces. For instance, to evaluate the APIs for the Cronos project, the experts received a list with 32 ``first tokens'' and a list with 73 ``second tokens'' automatically aggregated using SQL commands for each token. Finally, the experts analyzed the complete list (tokens + similarity suggestions) to pick one suggestion or decide using their experience. The whole process is illustrated in Figure \ref{fig:NLPExpert} and exemplified below. Table \ref{tab:APIclassification} shows the number of APIs evaluated by the experts in two rounds after the aggregations.} Therefore, instead of classifying the 441 APIs found in Cronos source code, they checked the NLP suggestions in the list of first and second tokens.

\begin{table}[ht]
 \centering
 \caption{Number of APIs classified per project }
 \label{tab:APIclassification}
\begin{tabular}{r|r|r|r|r|r}
\hline
\textbf{Project} &   \begin{tabular}[c]{@{}l@{}}\textbf{Total} \\ \textbf{APIs}\end{tabular} & \begin{tabular}[c]{@{}l@{}}\textbf{APIs} \\ \textbf{submited} \\ \textbf{to experts} \\ \textbf{1st token} \end{tabular} & \begin{tabular}[c]{@{}l@{}}\textbf{APIs} \\ \textbf{submited} \\ \textbf{to experts} \\ \textbf{2nd token} \end{tabular} & \begin{tabular}[c]{@{}l@{}} \textbf{APIs \%}\\ \textbf{1st round} \end{tabular}  & 
\begin{tabular}[c]{@{}l@{}}\textbf{APIs \%} \\ \textbf{1st + 2nd} \\ \textbf{round} \end{tabular}  \\
\hline
Audacity            & 1,478          &    562     & 106 & 38.0\% & 45.1\%\\
PowerToys           & 264           &    37      & 20 & 14.0\% & 21.6\% \\
Rmca                & 8,645          &    10      & 95 & 1.7\% & 2.3\% \\
Cronos              & 441           &    32      & 73 & 7.2\% & 23.8\%\\
JabRef              & 1,692          &    137     & 45 &  8.0\% & 10.8\% \\
\hline
\textbf{Total / avg}            & 12,520	        &    869	 & 339 & 6.9\% & 9.64\%\\
\hline
\end{tabular}
\end{table}

The process employed three experts (one author and two senior developers) and a card-sorting approach to manually accept or reject the suggestions for each token in the list. Each expert picked up one of the suggestions or chose a better API domain based on their experience. The experts could also check the list of full namespaces if they did not agree with the NLP suggestions. For example, considering the namespace ``com.oracle.xml.util.XMLUtil:'' for the first token, ``xml'', the similarity function suggested possible API-domain labels and a similarity value: Input and Output: 0.7, Error Handling: 0.69, Parser: 0.57. For the token ``util'', it suggested:  Utility: 0.9, Data Structure: 0.49. Therefore, the namespace ``com.oracle.xml.util.XMLUtil'' was classified as ``Utility.'' 
The majority of the APIs were classified using the first or second token. In a few cases ($<$ 10\%), the experts had to classify the full namespace. After classifying all the tokens, the experts conducted a second round to achieve consensus ($\sim$16 hours for all projects).

\responsetoreviewer{RevThreeCommentEight}{The project in Portuguese followed the same expert classification process employed in English projects. Indeed, the libraries declared in the Cronos project source code are written using English words and therefore did not harm the NLP categorization.}



We used these 31 categories (API-domains labels) for the 22,231 issues previously collected based on the presence of the corresponding APIs in the changed files. 
We used this annotated set to build our training and test sets for the multi-label classification models. 


\subsubsection{Phase 3 - Building the Multi-label Classifiers}
\label{sec:MultilabelClassification}

Since solving an issue may require multiple types of APIs, we applied a multi-label classification approach, which has been used in software engineering for purposes such as classifying questions in Stack Overflow (e.g.,~\cite{xia2013tag}) and detecting types of failures (e.g.,~\cite{feng2018empirical}) and code smells (e.g.,~\cite{guggulothu2020code}). To build the classifiers, we first needed to build the corpus and then run and evaluate the classifiers. 

\textbf{Corpus construction.}
\label{sec:DataExtraction}
The corpus construction comprised pre-processing, cleaning, diagnostics, and splitting into training and test datasets.  

\MyPara{Pre-processing:}
We built two distinct models---one that uses TF-IDF~\cite{ramos2003using} and another that uses BERT~\cite{ravichandiran2021getting}. 
These corpora include the issue title, body, and comment texts of the selected issues. 


Next, similar to other studies~\cite{ramos2003using,behl2014bug,vadlamani2020studying}, we applied TF-IDF, which is a technique for quantifying word importance in documents by assigning a weight to each word. After applying TF-IDF, we obtained a vector of TF-IDF scores for each issue's word. The vector length is the number of terms used to calculate the TF-IDF, and each term received the TF-IDF score. These TF-IDF scores are then passed to one of the selected classifiers (e.g., RandomForest) to label each issue. Each label receives a binary value (0 or 1), indicating whether the corresponding API domain is present in the issue.

For BERT, we created two separate CSV files: an input binary with expert API-domain labels paired with the issue corpus, as well as a list of the possible labels for the specific project. BERT directly labels the issue with the corpus text and labels list without the need for an additional classifier. 

We also evaluated the classifier's performance by combining in one dataset all the projects that use English vocabulary. Therefore, we also had to build a new composed ID (ID + project name) for all projects to guarantee uniqueness. For this experiment, after we created the new IDs, we merged the binaries of the project, including the classes missing for each project (RTTS does not have a Computer Graphics label, for example). We compared various algorithms to identify the best setup.

\MyPara{Cleaning:}
To build our classification models using TF-IDF, we converted each word in the corpus to lowercase and removed URLs, source code, numbers, and punctuation. We also removed stop-words and stemmed the words using the Python nltk package. We filtered out the issue and pull request templates\footnote{\url{http://bit.ly/NewToOSS}} since their repetitive structure introduced noise and were not consistently used among the issues.

We follow the work of \citet{TopicRecforSWRep} to process data for BERT. 
\responsetoreviewer{RevThreeCommentThirteen}{We tested BERT with a cleaned and uncleaned corpus. This was checked by comparing the F-measure, precision, and recall results from training with cleaned and uncleaned corpora. We ran three training trials with a 10-fold ShuffleSplit CV and determined that the unclean corpus consistently delivered higher metrics than any cleaning method (stemming, removing stopwords, etc.) The result is in line with \citet{TopicRecforSWRep} who showed that an unclean input corpus best maintained the context of words needed for BERT to determine their meaning and significance.} 

\MyPara{Diagnostics:}
\label{sec:Diagnostics}
Multi-label datasets are usually described by label cardinality and label density~\cite{MultilabelBook}. Label cardinality is the average number of labels per sample. Label density is the number of labels per sample divided by the total number of labels, averaged over the samples. For our dataset, the label cardinality is 8.19 and the density is 0.26. These values consider the 22,231 distinct issues and API-domain labels obtained after the previous section's pre-processing steps. Since our density can be considered high, the multi-label learning process or inference ability is not compromised~\cite{blanco2019multi}.

\MyPara{Training/Test Sets:}
\label{sec:TrainTest}
We split the data into training and test sets using the ShuffleSplit method~\cite{MultilabelBook}, which is a model selection technique that emulates cross-validation for multi-label classifiers. For example, in the JabRef project, we had 1,914 linked issues, and since one PR could be linked with more than one issue, we kept 1,648 entries that we randomly split into a training set with 80\% (1,318), 70\% (1,154), and 60\% (989) of the issues and a test set with the remaining 20\% (330 issues), 30\% (494), and 40\% (659). We ran each experiment ten times, using ten different training and test sets to match 10-fold cross-validation. To improve the balance of the data set, we ran the SMOTE algorithm for the multi-label approach~\cite{charte2015mlsmote}.

\textbf{Classifiers.}
\label{sec:Classsifier}
To create the classification models, we chose six classifiers that work with the multi-label approach and implemented different strategies to create learning models: Decision Tree, Random Forest (ensemble classifier), MLPC Classifier (neural network multilayer perceptron), MLkNN (multi-label lazy learning approach based on the traditional K-nearest neighbor algorithm)~\cite{zhang2007ml,MultilabelBook}, Logistic Regression, and BERT. We ran the \responsetoreviewer{RevThreeMinorEleven}{first five} classifiers using the Python sklearn package and tested several parameters. For the RandomForestClassifier, the best classifier, we kept the following parameters: \textit{criterion = `entropy', max\_depth = 50, min\_samples\_leaf = 1, min\_samples\_split =3, n\_estimators = 50}. 

The BERT model was built using the open-source python package, Fast-Bert~\cite{fastbert}, which builds on the Transformers~\cite{transformers} library for Pytorch. Before training the model, the optimal learning rate was computed using a lamb optimizer~\cite{You2020Large}. Finally, the model fit over 11 epochs and validated every epoch. This training and validation occurred for every fold in the ShuffleSplit 10-fold cross-validation. The BERT model was trained on an NVIDIA Tesla V100 GPU that is contained within a computing cluster. 
The choice of hardware is not critical so long as the target GPU has sufficient VRAM to train the BERT model. 

\MyPara{Classifiers Evaluation:}
\label{sec:Evaluation}
To evaluate the classifiers, we employed the following metrics (also calculated using the scikit-learn package):

\begin{itemize}

\item \textbf{Hamming loss} measures the fraction of wrong labels to the total number of labels.

\item \textbf{Precision} measures the proportion between the number of correctly predicted labels and the total number of predicted labels. 
 
\item \textbf{Recall} corresponds to the percentage of correctly predicted labels among all relevant labels. 
 
\item \textbf{F-measure} calculates the harmonic mean of precision and recall. F-measure is a weighted measure of how many relevant labels are predicted and how many of the predicted labels are relevant.
 
\end{itemize}

\begin{equation*}\tag{\roman{num_eq}}
   Precision = \frac{TP}{TP + FP}
\end{equation*}\stepcounter{num_eq}

\begin{equation*}\tag{\roman{num_eq}}
   Recall = \frac{TP}{TP + FN}
\end{equation*}\stepcounter{num_eq}

\begin{equation*}\tag{\roman{num_eq}}
   FMeasure = \frac{2TP}{2TP + FP + FN}
\end{equation*}\stepcounter{num_eq}

\responsetoreviewer{RevOneCommentOne}{The classic formulas to compute precision (i), recall (ii), and F-measure (iii) based on TP, TN, FP, and FN (true positives, true negatives, false positives, and false negatives) traditionally address single-label problems. An instance is considered correct or incorrect in single-label problems, while an instance may be partially correct in a multi-label evaluation; i.e., only a subset of the classes is correct for some instances. To address the multi-label classification problem, the literature \cite{tsoumakas2009mining} suggests adapting the aforementioned metrics as follows.


The metrics for each label can be calculated using different averaging strategies, as described in the following formulas. Let $TP_l$, $FP_l$, $TN_l$, and $FN_l$ be the number of true positives, false positives, true negatives, and false negatives returned by a binary evaluation effort $B(TP, TN, FP, FN)$ such as the binary relevance transformation for a label $l$ \cite{pereira2018correlation} and $q$ is the number of labels. The macro averaging \cite{tsoumakas2009mining} is the arithmetic mean of all the per-label metrics, while micro averaging \cite{tsoumakas2009mining} is the global average metric obtained by summing TP, FN, and FP. The averages are computed and used to calculate the precision, recall, and F-measure (i, ii, iii). \citet{santos2021can} used micro averaging to calculate the predictions' metrics. Thus, we kept it to compare with the previous study. The micro average favors the most populated classes \cite{sokolova2009systematic}.

\begin{equation}\tag{\roman{num_eq}}
    B_{macro} = \frac{1}{q} \sum^q_{l=1} B(TP_l, FP_l, TN_l, FN_l)
\end{equation}\stepcounter{num_eq}

\begin{equation}\tag{\roman{num_eq}}
    B_{micro} = B\left( \sum^q_{l=1} TP_l, \sum^q_{l=1} FP_l, \sum^q_{l=1} TN_l, \sum^q_{l=1} FN_l \right)
\end{equation}\stepcounter{num_eq}  

}

\textbf{Transfer Learning.}
\label{sec:TransferLearning}
Next, we investigate the behavior of the metrics when we use different sets to train and test the model. 
We combined four projects using English vocabulary using three projects for training and one for testing. For instance, we trained a dataset with JabRef, PowerToys, and Audacity to test using the RTTS project. Next, we substituted the test dataset with one in the training set until completing all possible combinations. 

\textbf{Data Analysis.}
\label{sec:DataAnalysis}
We used the aforementioned evaluation metrics, and the confusion matrix logged after each model's execution to evaluate the classifiers. We used the Mann-Whitney U test to compare the classifier metrics, followed by Cliff's delta effect size test. The Cliff's delta magnitude was assessed using the thresholds provided by \citet{Romano:2006}, i.e. $|$d$|$$<$0.147 ``negligible'', $|$d$|$$<$0.33 ``small'', $|$d$|$$<$0.474 ``medium'', otherwise ``large''. We considered p-value $<$ 0.05 as the limit to determine a statistical difference.



For the remainder of our analysis, we filtered out the API labels with no occurrence. 
``Cloud'' and ``Machine Learning'' did not appear in any issues/PR mined and, therefore, had no predictions. 
We also filtered out labels that appeared in more than 90\% of rows when running models for each project. Those could bias our predictions, since the classifier could always suggest them. 
For PowerToys, for example, the labels ``NLP'', ``Network'', ``DB'', ``Error Handling'', ``Language'', ``DevOps'', ``IO'', ``ML'', ``Security'', ``Cloud'', ``Event Handling'', ``CG'', ``Multimedia'', ``Thread'', ``Big Data'' and ``GIS'' had no occurrences and therefore were removed. The label ``Util'' was also removed because it surpassed the labels threshold (presented more than 90\% of the rows in the dataset). The ``Util'' label was the most present with 699 occurrences, followed by the 501 occurrences of ``App'' and 498 of ``UI''. The less representative set had ``Test'' (6), ``Logging'' (6), and ``i18n'' (4). PowerToys is a set of utility tools for Microsoft Windows. The high frequency of ``Util'' labels is expected.

\begin{figure*}[!hbt]
\centering
\hspace*{-2cm} 
\includegraphics[width=1.3\textwidth, trim= 50px 10px 50px 10px] 
{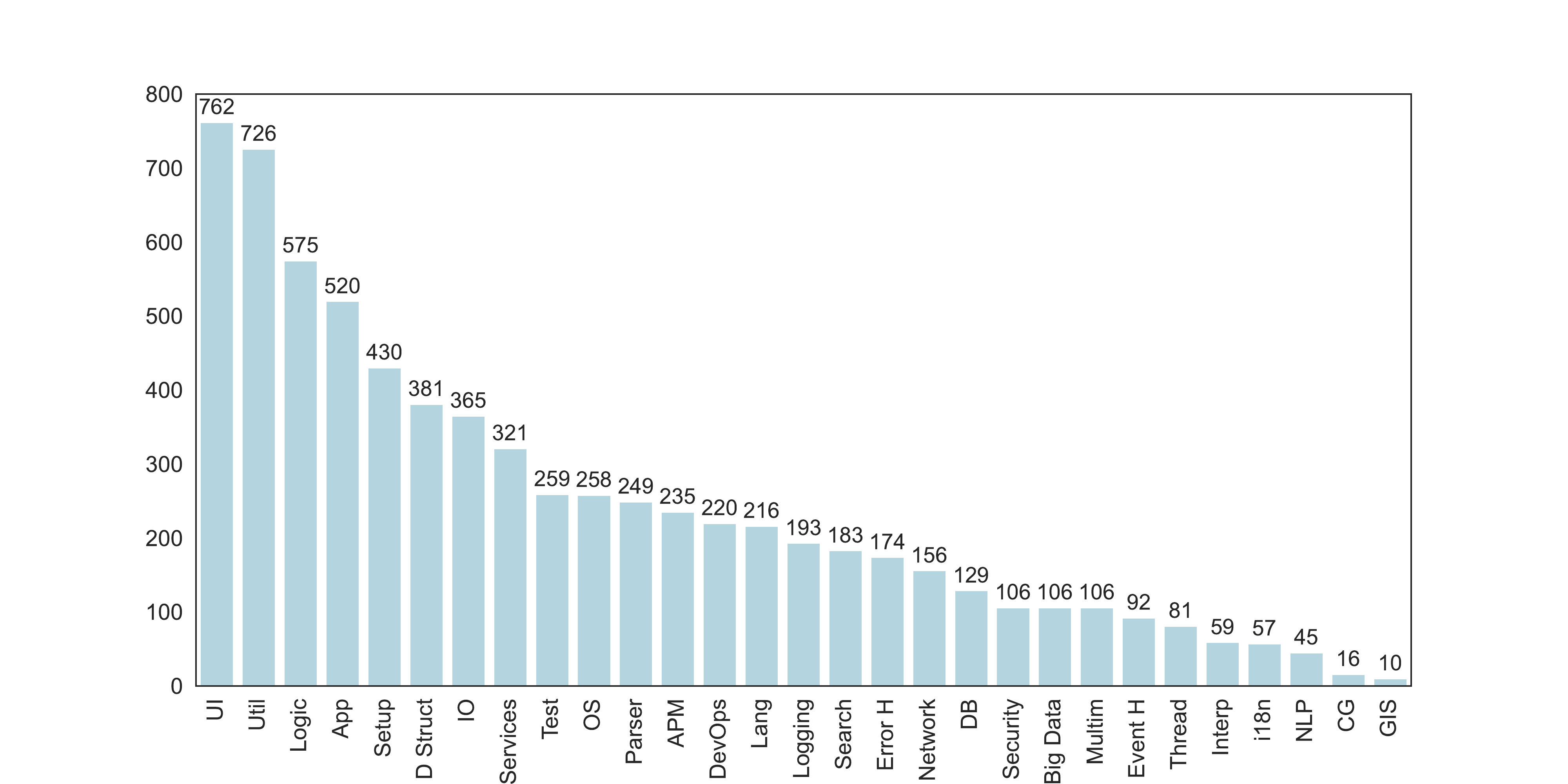}
\caption{Number of labels per type \responsetoreviewer{RevThreeCommentFivePartA}{from the model with all the datasets merged}}
\label{fig:labelFrequency}
\end{figure*}

The predictions using the dataset with all projects  considerably changed our distribution of labels. The most frequent Labels were ``UI'' with 762 occurrences, followed by ``Util'' with 726 and ``Logic'' with 575. The less frequent labels were: ``NLP'' (45), ``CG'' (16), and ``GIS'' (10). Despite some labels being popular and having been used for tagging many APIs by the experts, the lack of pull requests submitted that touched source codes with those APIs may explain their rareness. The lack of linked issues and pull requests that mention those labels can also cause the absence in the dataset. Finally, training all the datasets together helped to spread the labels' frequency, for instance: ``Util'' and ``Logic'' labels were dropped when training the JabRef project because they reached the threshold of 90\% of label predictions. When training using the dataset with all projects combined, those labels prevailed, staying below the 90\% threshold, and were used to tag the issues (Figure~\ref{fig:labelFrequency}). 

\begin{figure*}[!hbt]
\centering
\includegraphics[width=1\textwidth, trim= 50px 20px 50px 40px] {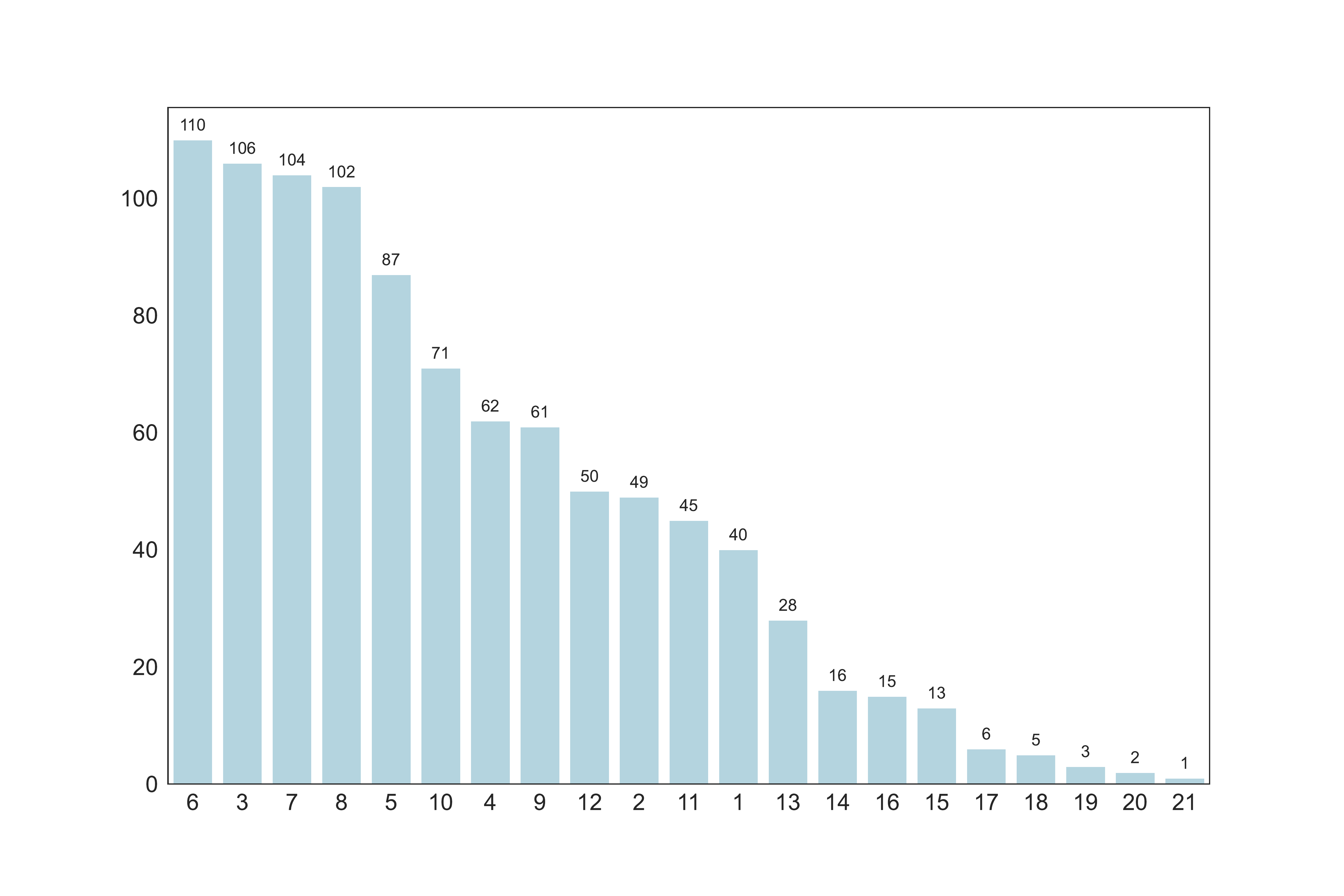}
\caption{Number of labels per issue \responsetoreviewer{RevThreeCommentFivePartB}{from the model with all the datasets merged}}
\label{fig:issueFrequency}
\end{figure*}

Finally, we checked the distribution of the number of labels per issue (Figure~\ref{fig:issueFrequency}). We found 110 issues with six labels, 106 issues with three labels, 104 issues with seven labels, and 102 issues with eight labels. Only 4.1\% (=40) of issues have one label, which confirms a multi-label classification problem (Figure~\ref{fig:issueFrequency}).

\subsection{Results}
\label{sec:RQ2-results} 


\textbf{\emph{RQ.2.1:}} 
To what extent can we automatically attribute API-domain labels to issues using data from the project?
\newline

\boldification{We compared the label predictions of many configurations, varying both the model algorithm, and the input data (title, body, comments).}

To predict the API-domains labels, we started by testing a simple corpus: only the issue \textsc{title} as input and the Random Forest (RF) algorithm, since it is insensitive to parameter settings~\cite{RandomForestShane} and has shown to yield good prediction results in software engineering studies~\cite{petkovic2016using,goel2017random,pushphavathi2014novel,satapathy2016early}. Then, we evaluated the corpus configuration alternatives, varying the input information: only \textsc{title} (T), only \textsc{body} (B), \textsc{title} and \textsc{body} (T+B), and \textsc{title}, \textsc{body}, and \textsc{comments} (T+B+C) comparing the average of all projects. To compare the different corpus configuration, we kept the Random Forest algorithm and used the Mann-Whitney U test with the Cliff's-delta effect size.

\boldification{We also tested alternative configurations using n-grams}

We also tested alternative configurations using n-grams. For each step, the best configuration was kept. Then, we used different machine learning algorithms and compared them to a dummy (random) classifier.



\boldification{For TF-IDF predictions, Title by itself had the worst performance. For BERT, it was the best one}

\begin{figure}
\centering
\includegraphics[width=1\textwidth, trim= 0px 25px 20px 40px]{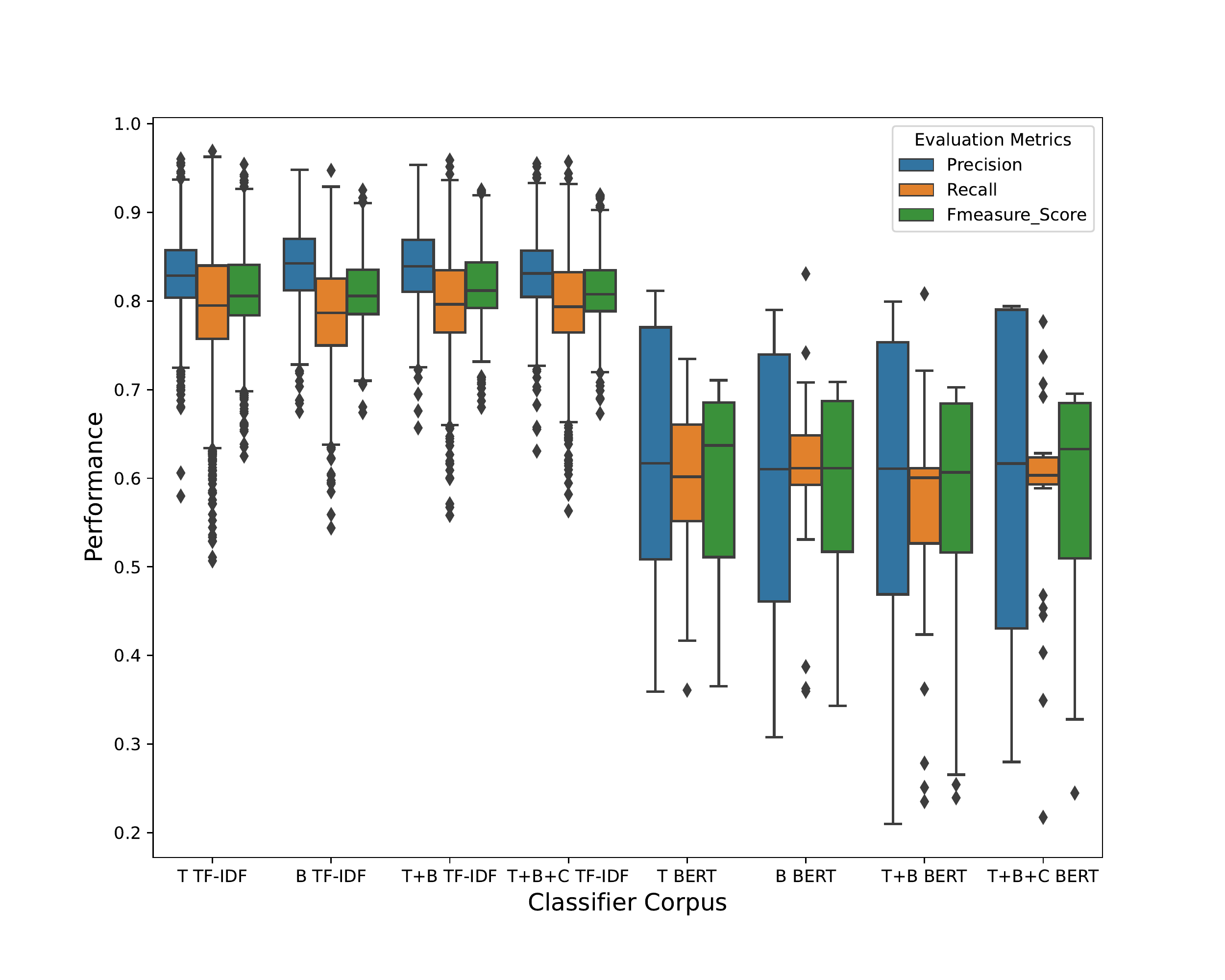}
\caption{Comparison between the corpus models inputted to TF-IDF and BERT. T=Title, B=Body, C=Comments}
\label{fig:H1H2H3Baseline-comparisonv3}
\end{figure}

As Figure~\ref{fig:H1H2H3Baseline-comparisonv3} and Table~\ref{tab:resultsTFIDF-BERT} (Appendix~\ref{sec:AppendixA}) show, when we tested different inputs and compared them to \textsc{Title} only, all alternative settings provided better results with TF-IDF. We observed improvements in terms of precision, recall, and F-measure from the previous study~\cite{santos2021can}. When using \textsc{body}, we reached a precision of 84\%, recall of 78.6\%, and F-Measure of 81.1\%. In contrast, while BERT had worse results, the model with the \textsc{Title} outperformed the other BERT models with 61.6\% precision. 

\begin{table}[ht!]
 \begin{center}
 \caption{Cliff's Delta for F-Measure and Precision: comparison of corpus model alternatives for TF-IDF and BERT. Title(T), Body(B) and Comments (C).}
 \label{tab:resultsH1H2H3cliffsTF-IDF-BERT}
 \begin{tabular}{l|l|r|l|r|l} 
 \hline
  \textbf{TF-IDF/BERT} & \textbf{Corpus} & \multicolumn{4}{c}{\textbf{Cliff's delta}} \\
 \cline{3-6}
 \textbf{ } & \textbf{Comparison} & \multicolumn{2}{c|}{\textbf{F-measure}} & \multicolumn{2}{c}{\textbf{Precision}}\\
 \hline
TF-IDF & T versus B & -0.005 & negligible & -0.15 & small***\\
TF-IDF & T versus T+B & -0.10 & negligible*** & -0.12 & negligible***\\
TF-IDF & T versus T+B+C & -0.03 & negligible*** & -0.01 & negligible \\
TF-IDF & B versus T+B & 0.10 & negligible*** & 0.02 & negligible \\
TF-IDF & B versus T+B+C & -0.02 & negligible & 0.14 & negligible*** \\
TF-IDF & T+B versus T+B+C & 0.07 & negligible*** & 0.11 & negligible***\\

BERT & T versus B & 0.07 & negligible & 0.11 & negligible \\
BERT & T versus T+B & 0.13 & negligible & 0.03 & negligible \\
BERT & T versus T+B+C & 0.03 & negligible & 0.09 & negligible \\
BERT & B versus T+B & 0.10 & negligible & -0.04 & negligible \\
BERT & B versus T+B+C & -0.006 & negligible & 0.08 & negligible \\
BERT & T+B versus T+B+C & -0.09 & negligible & -0.01 & negligible\\
\hline
\multicolumn{5}{l}{\scriptsize\textit{* p $\leq$ 0.05;
** p $\leq$ 0.01; 
*** p $\leq$ 0.001}}
\\

 \end{tabular}
 \end{center}
\end{table}

\boldification{The differences in the results were statistically significant in many TF-IDF comparisons, but they were not in BERT comparisons}

For TF-IDF, we found statistical differences comparing the results using \textsc{title} only and all the three other corpus configurations: F-measure (p-value $\leq$ 0.001 when comparing with \textsc{title}+\textsc{body} or \textsc{title}+\textsc{body}+\textsc{comments}, Mann-Whitney U test) and precision (p-value $\leq$ 0.001 when comparing with \textsc{body} or \textsc{title}+\textsc{body}, Mann-Whitney U test), both with negligible effect size when comparing the precision from \textsc{title} and \textsc{body}. The corpus configured with \textsc{body} performed better than all others in terms of precision, followed closer by the one set up with \textsc{title}+\textsc{body}, which performed better in recall and F-measure. However, the results suggest that using only the \textsc{body} would provide good enough outcomes since there was a negligible effect size compared to the other two configurations---using \textsc{title} and/or \textsc{comments} in addition to the \textsc{body}---achieving similar results with less effort. 
Table~\ref{tab:resultsH1H2H3cliffsTF-IDF-BERT} shows the Cliff's-delta comparison between each pair of corpus configurations, and Figure~\ref{fig:H1H2H3Baseline-comparisonv3} shows the box plots confirming the similar results carried out by the three diverse setups. For BERT, all the models 
had the same distribution in precision and F-measure.

Next, we investigated the use of bigrams, trigrams, and quadrigrams, comparing the results to the use of unigrams. We used the corpus with only the issue \textsc{body} for this analysis, since this configuration was chosen in the previous step. Table~\ref{tab:resultsGrams} (Appendix~\ref{sec:AppendixA}) and Figure~\ref{tab:resultsH5cliffs} present how the algorithms perform for each n-gram configuration. While the unigram configuration has a slightly better F-measure, the quadrigram has slightly better precision. However, their differences in the precision have a negligible effect size, and their differences in F-measure have a small effect size. Additionally, the unigram uses less computational effort and memory~\cite{van2016efficient}. Hence, we kept the unigram as the best option.




\begin{table}[ht!]
 \begin{center}
 \caption{Cliff's Delta for F-Measure and precision: Comparison between n-grams models}
 \label{tab:resultsH5cliffs}
 \begin{tabular}{l|r|l|r|l} 
 \hline
 \textbf{n-Grams} & \multicolumn{4}{c}{\textbf{Cliff's delta}} \\
 \cline{2-5}
 \textbf{Comparison} & \multicolumn{2}{c|}{\textbf{F-measure}} & \multicolumn{2}{c}{\textbf{Precision}}\\
 \hline

1 versus 2 & 0.09 & negligible*** & -0.02 & negligible**\\
1 versus 3 & 0.11 & negligible*** & -0.01 & negligible \\
1 versus 4 & 0.15 & small*** & -0.06 & negligible \\
2 versus 3 & 0.02 & negligible & 0.01 & negligible*** \\
2 versus 4 & 0.06 & negligible*** & -0.04 & negligible** \\
3 versus 4 & 0.04 & negligible** & -0.05 & negligible***\\
\hline
\multicolumn{5}{l}{\scriptsize\textit{* p $\leq$ 0.05;
** p $\leq$ 0.01; 
*** p $\leq$ 0.001}}
\\
 \end{tabular}
 \end{center}
\end{table}

\begin{figure}[!hbt]
\centering
\includegraphics[width=1\textwidth, trim= 10px 25px 20px 20px]{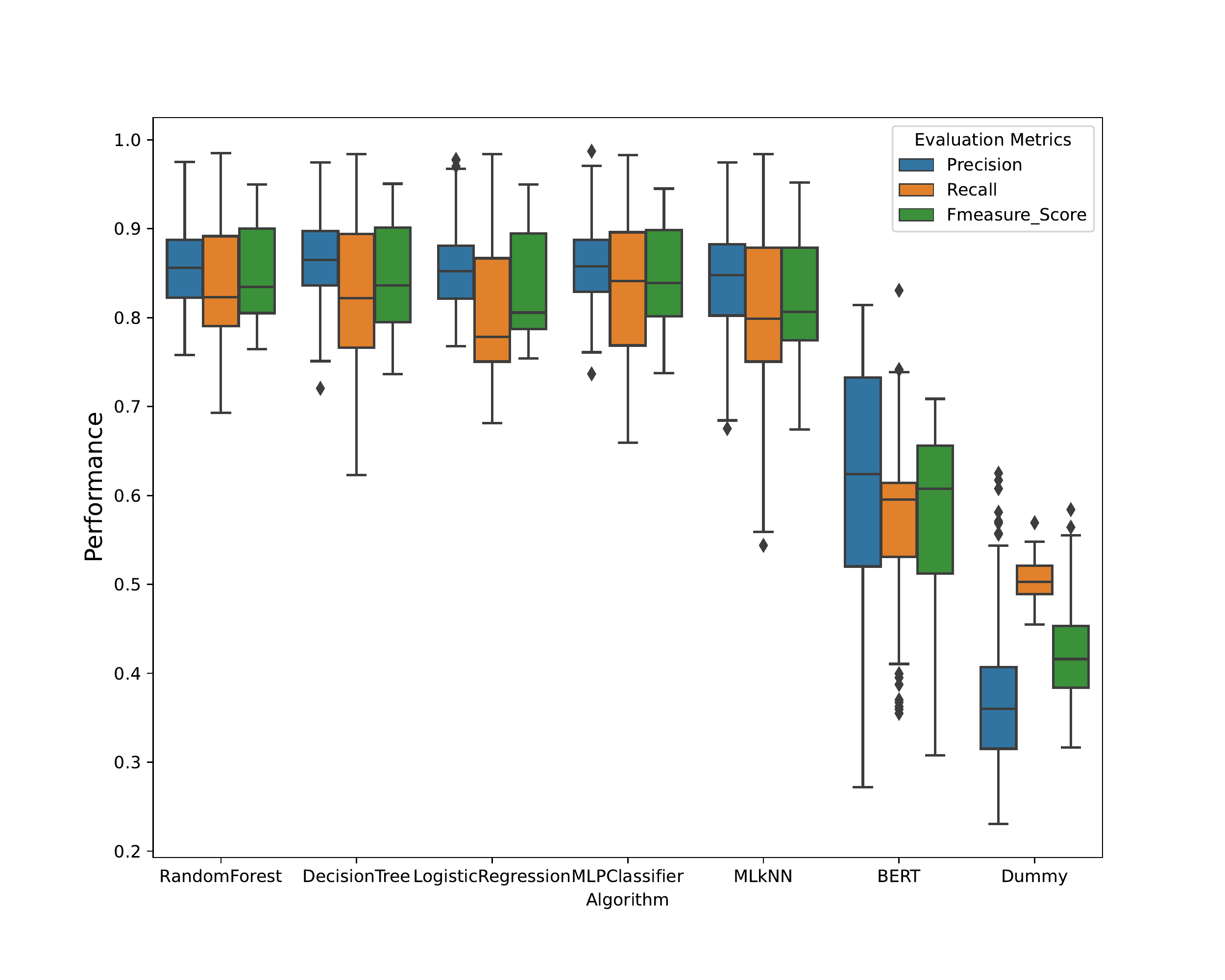}
\caption{Performance comparison between the machine learning algorithms}
\label{fig:baselineH6}
\end{figure}



\begin{table}[hbt!]
 \begin{center}
 \caption{Cliff's Delta for F-Measure and precision: Comparison between machine learning algorithms}
 \label{tab:resultsH6cliffs}
 \begin{tabular}{l|r|l|r|l} 
 \hline
\textbf{Algorithms} & \multicolumn{4}{c}{\textbf{Cliff's delta}} \\
 \cline{2-5}
\textbf{Comparison} & \multicolumn{2}{c|}{\textbf{F-measure}} & \multicolumn{2}{c}{\textbf{Precision}}\\
 \hline
RF versus LR & 0.27 & small*** & 0.06 & negligible*\\
RF versus MLPC & 0.02 & negligible & 0.009 & negligible \\
RF versus DT & 0.06 & negligible* & 0.09 & negligible*** \\
RF versus MlkNN & 0.28 & small*** & 0.13 & negligible*** \\
RF versus BERT & 1.0 & large*** & 1.0 & large*** \\
LR versus MLPC & -0.21 & small*** & -0.07 & negligible* \\
LR versus DT & -0.15 & small*** & -0.15 & small*** \\
LR versus MlkNN & 0.07 & negligible* & 0.08 & negligible* \\
LR versus BERT & 1.0 & large*** & 1.0 & large*** \\
MPLC versus DT & 0.03 & negligible & -0.08 & negligible*** \\
MPLC vs. MlkNN & 0.24 & small*** & 0.13 & negligible*** \\
MLPC versus BERT & 1.0 & large*** & 1.0 & large*** \\
MlkNN versus DT & -0.19 & small*** & -0.20 & small*** \\
MlkNN versus BERT*** & 1.0 & large & 1.0 & large*** \\
DT versus BERT*** & 1.0 & large & 1.0 & large*** \\
RF versus Dummy & 1.0 & large*** & 0.50 & large*** \\

\hline
\multicolumn{5}{l}{\scriptsize\textit{* p $\leq$ 0.05;
** p $\leq$ 0.01; 
*** p $\leq$ 0.001}}
\\
 \end{tabular}
 \end{center}
\end{table}

\boldification{We also investigate the influence of ML classifiers using the body with unigrams as corupus}

To investigate the influence of the machine learning (ML) classifier, we compared several options using the \textsc{body} with unigrams as a corpus. The options included: Random Forest (RF), Neural Network Multilayer Perceptron (MLPC), Decision Tree (DT), LR, MlKNN, BERT, and a Dummy Classifier with strategy ``uniform.'' Dummy or random classifiers are often used as a baseline~\cite{saito2015precision, flach2015precision}. We used the implementation from the Python package scikit-learn\footnote{\url{https://scikit-learn.org/}}. 
Figure~\ref{fig:baselineH6} and Table~\ref{tab:resultsAlg} (Appendix~\ref{sec:AppendixA}) show the comparison among the algorithms, and Table~\ref{tab:resultsH6cliffs} presents the pair-wise statistical results comparing F-measure and precision using Cliff's delta. 


\boldification{Random forest was the best one}

Random Forest (RF) was the best model when compared to Decision Tree (DT), Logistic Regression (LR), Neural Network Multilayer Perceptron (MLPC), MlKNN algorithms, and BERT. Random Forest outperformed these five algorithms with negligible/small effect sizes considering F-measure and precision. Compared to BERT and the Dummy Classifier, the effect size was large. \responsetoreviewer{RevThreeCommentNine}{The observed difference among some algorithms are fairly small and therefore might vary according to project corpus properties.}  

\boldification{The classification metrics indicate that the random forest classifier is suitable for predicting labels}

The results showed the classifier is suitable for predicting labels in projects written in different programming languages (C++, C\#, and Java), with issues with vocabulary in English and Portuguese.

\MyBox{\textbf{\emph{RQ.2.1 Summary.}} It is possible to individually predict the API-domain labels for each project with a precision of 0.864, recall of 0.786, and F-measure of 0.811 
using the Random Forest algorithm, \textsc{body} as the corpus, and unigrams.}

\textbf{\emph{RQ.2.2:}} 
To what extent can we automatically attribute API-domain labels to issues using data from other projects?
\newline

\boldification{We merged the datasets to check how much labels from a project can help another, and found that the performance of metrics decreases, but BERT was less impacted than the other models. Nonetheless, Random Forest still had the best precision}

\begin{figure}
\centering
\includegraphics[width=1\textwidth,trim= 10px 25px 20px 20px]{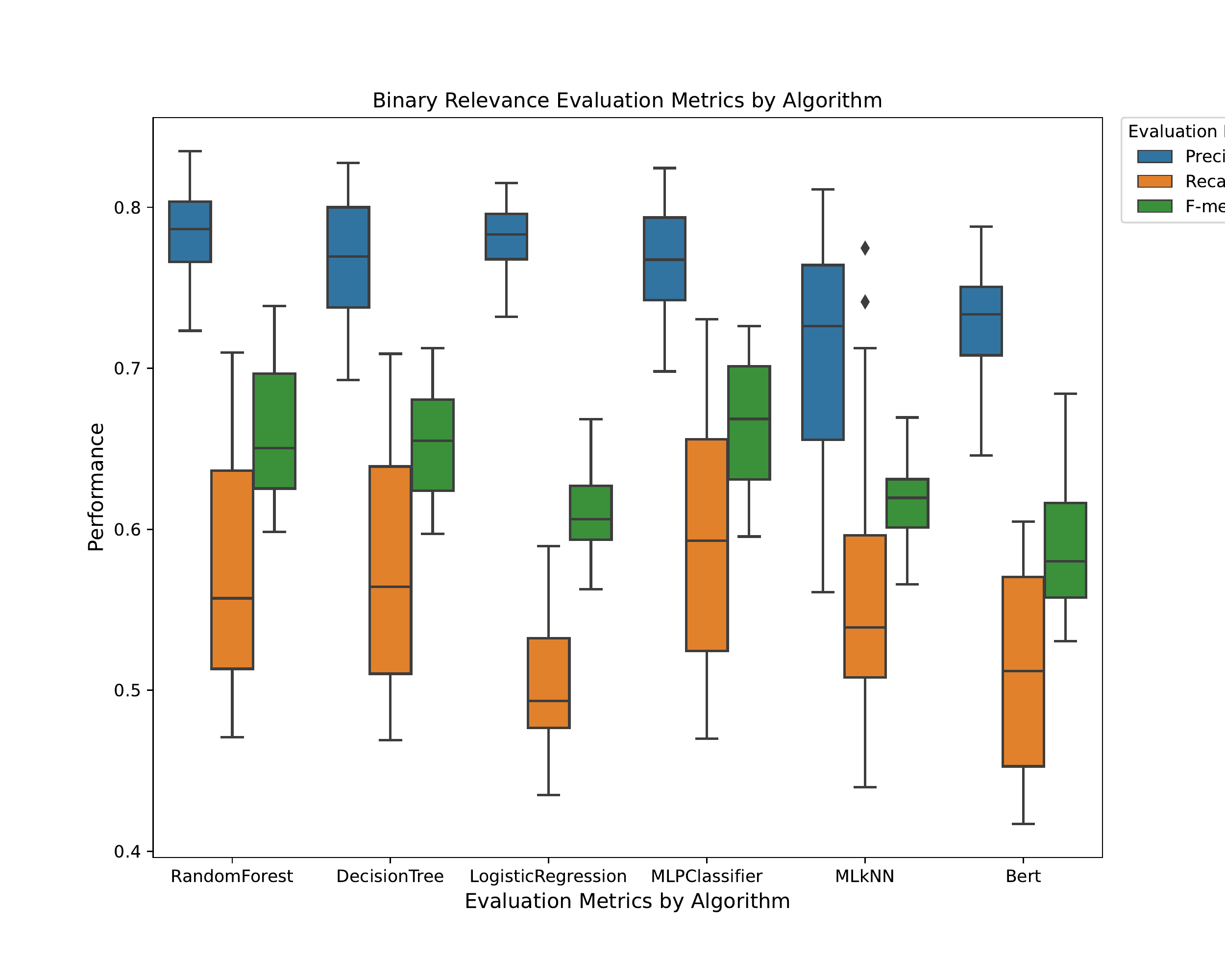}
\caption{Performance comparison between machine learning algorithms using the dataset with all projects - Vocabulary: EN}
\label{fig:baselineH7}
\end{figure}

Next, we merged the datasets that use English vocabulary (RTTS, JabRef, Audacity, and PowerToys), predicting the API-domain labels for all the projects. Removing the project with Portuguese vocabulary was necessary since the BERT model was trained with English vocabulary. The predictions were carried out with \textsc{body} as the corpus (and unigrams for the TF-IDF). Figure~\ref{fig:baselineH7} shows the performance obtained with diverse algorithms. RF still had the best precision while the MLPC had the best F-measure; BERT had better precision than MLkNN and better recall than Logistic Regression. BERT was less impacted by the loss of metrics when predicting the API-domain labels with the all-projects combined dataset (Table~\ref{tab:resultsAllTogether} - Appendix~\ref{sec:AppendixA}).

\MyBox{\textbf{\emph{RQ.2.2 Summary.}} Predicting using a dataset with all English-language projects combined decreased the precision by 9.15\% using Random Forest and increased the precision using BERT by 20.63\%.}

\textbf{\emph{RQ.2.3:}} 
To what extent can we automatically attribute API-domain labels to issues using transfer learning?
\newline
\begin{table}[ht]
 \caption{Overall performance from models created to evaluate the transfer learning. }
 \label{tab:transfer}

\begin{tabular}{l|l|l|l|l}
\hline
\textbf{Training}           & \textbf{Test} & \textbf{P} & \textbf{R} & \textbf{F} \\ \hline
RTTS, Audacity, PowerToys   & JabRef        & 0.305      & 0.294      & 0.299      \\
JabRef, Audacity, PowerToys & RTTS          & 0.713      & 0.175      & 0.281      \\
JabRef, RTTS, PowerToys     & Audacity      & 0.688      & 0.284      & 0.402      \\
JabRef, RTTS, Audacity      & PowerToys     & 0.296      & 0.525      & 0.379      \\ \hline
JabRef, Audacity, PowerToys & RTTS* & 0.718 & 0.272 & 0.394 \\
\hline
\multicolumn{5}{l}{\scriptsize\textit{* RTTS - labels most important to the users (Section \ref{sec:RQ3-results})}}
\end{tabular}
\end{table}

\boldification{We also evaluated how the prediction of the projects transfers to a new one. We found big discripancies} 

Finally, Table~\ref{tab:transfer} shows the results for all combinations tested with transfer learning. The results had a significant range in precision and recall varying from 0.713 to 0.296 in precision predicting RTTS and PowerToys, respectively, and from 0.525 to 0.175 in recall predicting Audacity and RTTS, respectively. 

\boldification{By restricting the prediction to labels evaluated (curated?) by developers, we found a increase in precision, recall, and f-measure}

\responsetoreviewer{RevOneCommentFourPartA}{Additionally, we ran a transfer learning experiment targeting the RTTS project labels evaluated by developers (Section \ref{sec:RQ3-results}). We dropped all labels with fewer than three evaluations and up to 50\% of ``Not Important'' evaluations (see Figure \ref{fig:contributorsassesmenRTTSCronos}). Therefore, in the RTTS project, the labels that persisted are: ``Network'', ``Logging'', ``Setup'', ``Micro/services'', and ``UI''. Since Audacity, JabRef, and PowerToys projects were not evaluated by developers (Section \ref{sec:RQ3}), they were not included in this experiment. We observed a small increase in precision (0.713 to 0.718) and a significant increase in recall (9.7\% - 0.175 to 0.272) and F-measure (11.3\% - 0.281 to 0.394) - Table \ref{tab:transfer}}. 



\MyBox{\textbf{\emph{RQ.2.3 Summary.}} Transferring learning with diverse configurations considering source and target projects decreased the metrics from 15.12\% to 64.74\% and the recall from 33.21\% to 77.74\%, depending on the sources and target project. \responsetoreviewer{RevOneCommentFourPartB}{Evaluating the transfer learning concerning only the API-domain labels evaluated as important by the developer who solved the issues improved the recall by 9.7\% and F-measure by 11.3\%.}}


\section{RQ3 - Evaluating the API-domain labels with developers}
\label{sec:RQ3}


Considering human input is very relevant in machine learning studies, we labeled some issues and presented them to developers that solved the same issues previously to receive feedback about how useful the API-domain labels could be if available at the time they worked on the issues.

\subsection{Method}
\label{sec:RQ3Method}

To answer RQ3, we use the Random Forest algorithm, issue description \textsc{body} as the corpus, and unigrams (the best configuration we found in RQ.2) to generate labels for the issues. 




\subsubsection{Labels generation} 
We predicted labels for 91 issues (PowerToys = 21, Audacity = 18, Cronos = 24, and RTTS = 28). The predictions covered all the 29 proposed API-domain labels (Cloud and ML do not have samples in our projects). We selected the most recently closed issues from the projects to get better chances of finding the developer who fixed the issues and they recall the problem solved. However, some issues had to be discarded when the contributor who solved them was not working for the enterprise anymore or when the OSS contributors did not answer our contact (Section \ref{sec:ContributorsAssessment}). The use of the most recent issues and the availability of the participants created an unbalanced set of labels for evaluation and we use our best effort to include the most representative set of API-domain labels possible in the empirical experiment. 

\subsubsection{Contributors assessment} \label{sec:ContributorsAssessment} In this step, we recruited 20 participants (PowerToys (1), Cronos (13), and RTTS (6)). 
To recruit participants from those projects we sent emails to maintainers from PowerToys and Audacity and contacted development managers from Cronos and RTTS. We asked participants from those projects to evaluate if the labels represent the skills needed to solve the issues and could help newcomers or experienced developers who want to choose an issue.   
\responsetoreviewer{RevThreeCommentTenPartC}{All of the participants were experts in their project and were asked to evaluate the issues to which they contributed in the past. Indeed, the number of issues evaluated by participants varied according to their past contributions. Each issue was evaluated by only one participant. The participants received a gift card as a token of appreciation for their participation.}

We asked the following questions:

\begin{itemize}

 \item How important do you consider having these labels on the issue to help new contributors identify the skills needed to solve them? (Evaluate each label) 
 (Likert: Very Important, Important, Moderately Important, Slightly Important, Not Important)
 \item Why?
 \item What labels are missing?
\end{itemize}

\subsubsection{Analysis} Based on the data gathered in the contributors' assessment, we performed a quantitative analysis to assess the generated labels. To analyze the open questions in which the contributors could explain their opinions about the labels generated, we employed open coding and axial coding procedures~\cite{strauss1998basics}.

\subsection{Results}
\label{sec:RQ3-results} 

\boldification{We received feedback about 26 issues out of the 91 predicted ones -- with 16 different domain labels. We discuss the feedbacks from Cronos and RTTS below}

From the 91 issues predicted, we received 29.67\% feedback (26 issues and 16 different API-domain labels). We did not receive feedback from the Audacity contributors. PowerToys had only one contributor who evaluated only three issues encompassing only four labels. Due to insufficient data, we removed this project from the results.

\boldification{In Cronos, labels DevOps, UI, DB, Lang, and Security were classified as important. Lables APM, Setup, NLP, and IO were classified as not important}

\begin{figure}
\centering
\includegraphics[width=1\textwidth]{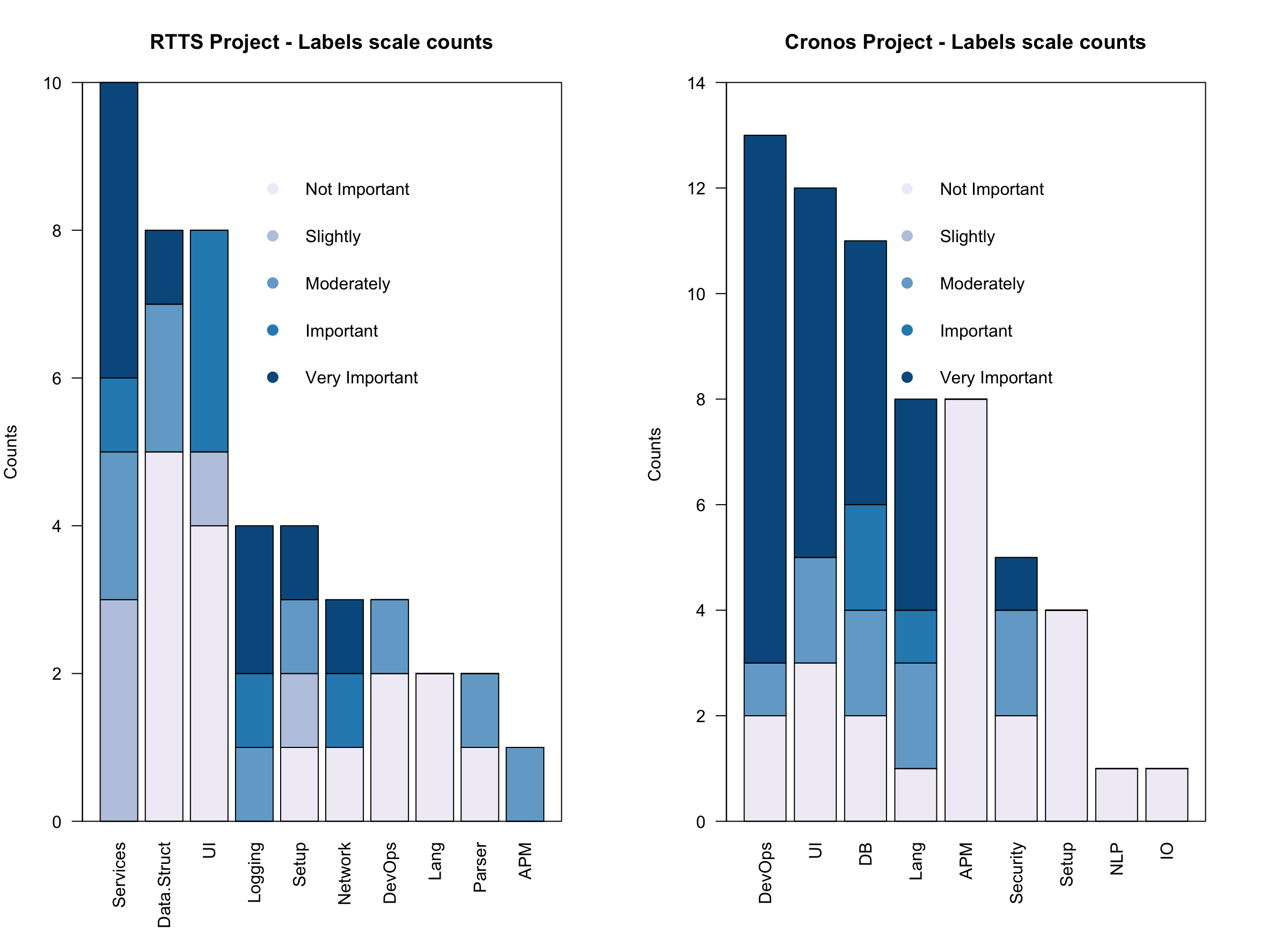}
\caption{Labels assessment by project}
\label{fig:contributorsassesmenRTTSCronos}
\end{figure}

\textbf{Cronos.} 
\label{sec:cronos}
\responsetoreviewer{RevThreeCommentTenPartA}{A total of 13 contributors assessed the generated labels.}
Based on the results (Figure~\ref{fig:contributorsassesmenRTTSCronos}), the contributors described 5 labels (i.e., DevOps, UI, DB, Lang, and Security) as very important or important 
and APM, Setup, NLP, and IO labels unimportant. DevOps, UI, DB, and Lang were highly rated as important, with many ``Very important'' and ``Important'' evaluations. \responsetoreviewer{RevTwoCommentNineteenPartA}{\responseinner{Not all the participants justified their response, but among the reasons} those contributors mentioned that ``\textit{It was a simple UI issue.}'' (P15) indicating the success of the ``UI'' prediction. Another developer mentioned ``\textit{This issue also required database, logic and lang skills.}'' (P6). This issue was tagged with ``UI'' and ``DevOps'' (evaluated as ``Very important'') but the developer missed some skills. \responseinner{Related to the missing labels, some contributors mentioned that ``\textit{The issue is related to a restriction. It requires UI skills and DevOps skills (not included in the predictions. [...]
But it also requires other skills.}'' (P18). Another contributor also missed some labels and mentioned ``\textit{This issue also required database, logic, and language skills.}'' (P06).}}


\boldification{In RTTS, labels Services, UI, Logging, Setup, Network, and Data Structure were classified as important. Labels Lang, Parser, DevOps, UI, and Data Structure were classified as not important}

\textbf{RTTS.} 
\label{sec:RTTS}
\responsetoreviewer{RevThreeCommentTenPartB}{Concerning the RTTS project, five contributors assessed the labels generated for the issues they had solved; in this scenario, we have contributors evaluating from 1 to 3 issues each.}
Our findings (Figure~\ref{fig:contributorsassesmenRTTSCronos}) highlight that the following labels were classified by contributors as important to very important: Services, UI, Logging, Setup, Network, and Data Structure. \responsetoreviewer{RevTwoCommentNineteenPartB}{\responseinner{Among the reasons contributors highlighted, we can observe the positive feedback} as mentioned in: ``\textit{I can totally agree with the labels for this, as to find the problem and apply the solution all the skills are necessary.}'' (P20). Moreover, contributors classified the following labels as not important or slightly important: Lang, Parser, DevOps, UI, and Data Structure. Some contributors mentioned that: ``\textit{I partially agree, some of the labels could give an initial point of view to the reported issue, but some are not related, like Language, Data structure and Setup}'' (P2). \responseinner{Some contributors reported missing some labels according to what was mentioned: ``\textit{Logging skills would be necessary to troubleshoot the issue and get the relevant information from the application, while service skills (knowledge about how service discovery and the service registry in the system works) would be necessary to find that the service version of the requested service didn't match what was registered. As for Network, it could have been useful to be able to determine that this issue was not caused by some error/faulty response from the requested service, but in this case, the log stated explicitly that the requested service did not exist. I don't find that the other labels/skills apply to this issue.}'' (P04).}}






\MyBox{\textbf{\emph{RQ.3. Summary.}} Our findings suggest the labels would be useful to help identify the skills needed to solve them. 
The efficiency of labels generated differs by project, for 
Cronos we had 61.9\% of the labels evaluated in the range from slightly to Very important and 64.4\% in the RTTS project in the same range.} 






\section{Discussion}\label{sec:discussion}

This section discusses our results and future work. 


\boldification{Developers like to see multiple characteristics on labels, but some characteristics are ambiguous}

\textbf{Do developers have a well-defined preference about labels?}
The feedback shared by study participants in Section~\ref{sec:RQ1} 
showed us the importance of the different types of labels to ease the issue selection process. However, the developers expressed preferences about different types of labels, and some preferences are ambiguous. For instance, P32 indicated “The technology used” when we asked what kind of labels they want to see in the issues. 
The technology could refer to a ``programming language,'' or an ``API.'' While both classifications could be used, we would prefer to define the technology as API because it is more specific than a programming language or even a framework that can encompass many libraries. A similar situation emerged with ``priority'' (P12). 
The ``priority'' could be restricted only to ``low'' or ``high,'' or could it include other aspects like the ``impact on operations'', as suggested by P15.  

\boldification{We can group the labels into two types: general and management. In this work, we focus on general API-domain labels, but future work should focus on finding a balance}

We can group the kind of labels in technology (technology, API, programming language, database, framework, and architecture layer) and management (type, priority, status, difficulty level, GitHub info).
Management labels are more often used in issue trackers. In this work, we propose to add to the issues a kind of technological label, the API-domain labels, which we claim are a proxy for the skills needed to solve an issue. 
Nonetheless, one should avoid overloading the issue trackers with too many labels. Future research can investigate the right balance of offering labels without creating a visual overhead for the contributor.

\boldification{For Newcomers, API-domain labels are more relevant than other types of general labels and slightly more relevant than management labels}

\textbf{Are API-domain labels relevant?} 
Our findings show that participants considered API-domain labels relevant in selecting issues. More specifically, newcomers to the projects considered API-domain labels more relevant than other general labels that describe the components and slightly more favored than management labels describing the type of issue. This suggests that a higher-level understanding of the API domain is more relevant than deeper information about the specific component in the project. 

\boldification{Experienced coders found the API-domain labels more useful than novices}

When controlling for issue type and component, API-domain labels were considered more relevant for experienced coders than novices (or students). This suggests that novices may need more help than ``just'' the technology for which they need skills. Our results also show that novices could be helped if the issues provide additional details about the complexity levels, how much knowledge about the particular APIs is needed, the required/recommended academic courses needed for the skill level, estimated time to completion, contact for help, etc. 

\boldification{A newcomer is not necessarily a novice. Experience from previous projects can help the onboarding and make API-labels more relevant}

\responsetoreviewer{RevTwoCommentTwentyOne}{Although each contributor is a newcomer when they move to a new project, previous experience counts when the new project shares technology with the previous projects. As opposed to experienced newcomers, who may transfer knowledge from previous projects and jump directly to the issue solution, novice newcomers spend more time understanding the project structure, the underlying technology, and how to set up the environment \cite{santos2022how} which might suggest why practitioners from the industry and experienced participants selected more API-domain labels than students and novices. Perhaps the API granularity is deeper than what the novices are looking for. Future research may consider the appropriate technical information to assist novice newcomers.}  
%
%

\boldification{Newcomers also indicate that they find skill information in the title, body, and comments of issues -- which can be organized with templates.}

\responsetoreviewer{RevTwoCommentTwentyThreePartB}{\textbf{In addition to API-domain labels, what issue characteristics are relevant to identify skills in issues?} 
In addition to labels, new contributors mentioned the \textsc{title}, \textsc{body}, and \textsc{comments} as sources of information to identify the necessary skills to work on the issues. Such elements can be structured with issue templates or written in an ad-hoc manner. \citet{santos2022how} asked maintainers to suggest community strategies to help newcomers find a suitable issue. Among the identified strategies, maintainers suggested 15 diverse ways of labeling the issues (e.g., labeling with skills, knowledge area, programming languages, libraries, and others) and several ways of organizing the issues, which include creating templates.

\boldification{However, some issues (and templates) do not have this information and the 5W2H analysis can help inspect their completeness}

While these other issue elements may indicate the skills and other characteristics of the issues that are not on the labels, some issues -- and existing templates -- are incomplete, lacking important information for contributors. The 5W2H analysis we applied in this paper can help us to \responseinner{holistically} understand what should be written in issues by \responseinner{covering the seven dimensions of information - who, why, when, what, where, how to solve and how big is the issue}. \responseinner{Future work can use the 5W2H questions to inspect the completeness of existing templates in terms of covering the seven dimensions of information. 

\boldification{While templates are useful for human contributors, we removed their repeated sentences in our NLP pipeline processing.}

Despite the importance of issue templates, we removed template sentences in an effort to clean repeated text to be ingested by the data processing pipeline. 
For example, one template sentence is ``Steps to reproduce.'' Since this fixed text appears in many issues (regardless of their categories) and the templates had changed over time, we decided to remove it before processing the issue corpus. This removal only affects the trained model, and we still should use the results of the 5W2H analysis to create a human-oriented template able to point new contributors to information relevant to them.}}

\textbf{What are the effects of the corpus characteristics on the labels' classifications?} 
Observing the reported results (TF-IDF) for different corpora used as input, we noticed that the model created using only the issue body performed similarly to the models using the issue title, body, and comments, and better than the model using only the title. By inspecting the results, we noticed that by adding more words to create the model, the matrix of features becomes sparse and does not improve the classifier's performance. 

\begin{figure*}
\centering
\hspace*{-2cm}  
\includegraphics[width=1.5\textwidth] {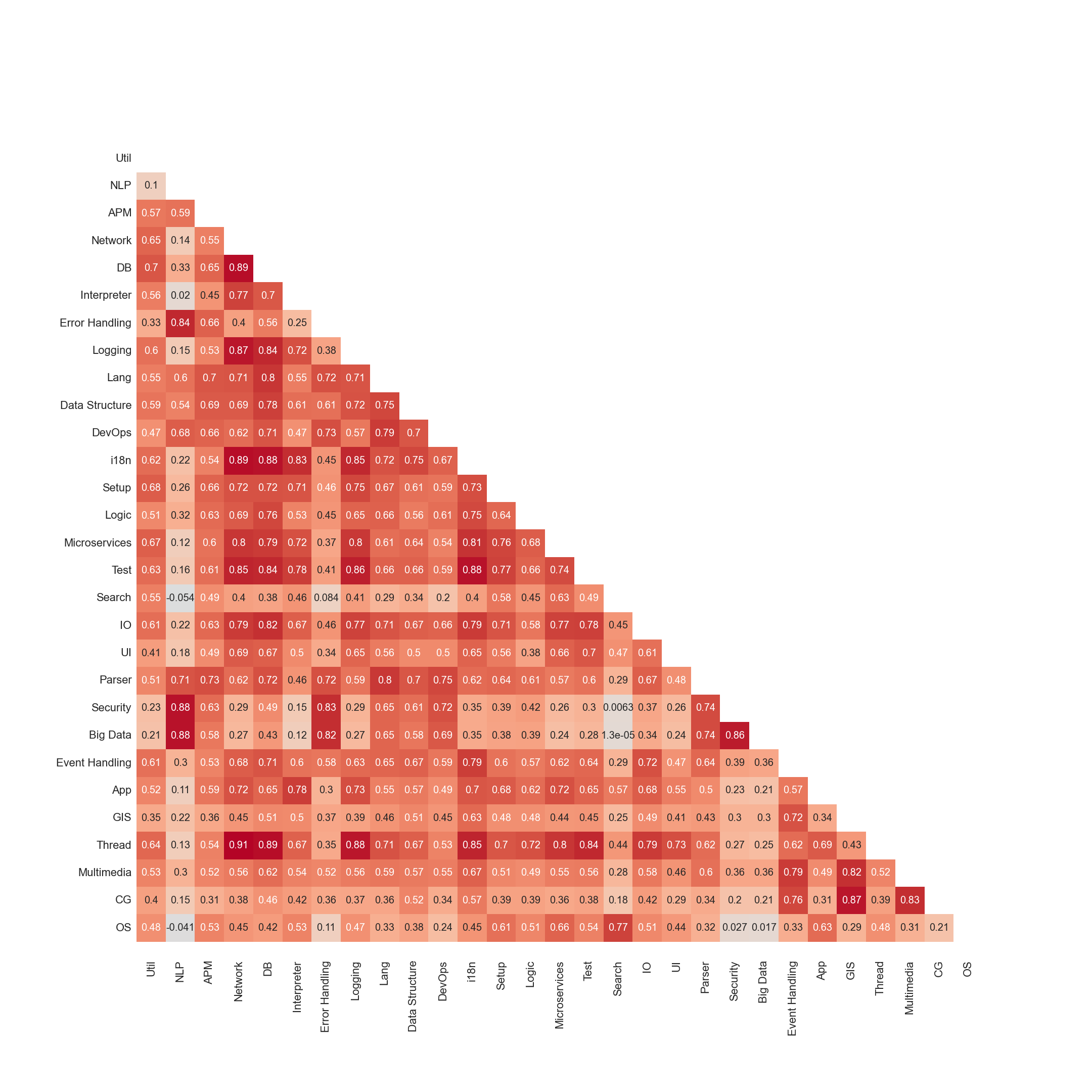}
\caption{Heat Map - Label correlation in the dataset with all projects combined. \responsetoreviewer{RevThreeMinorFourteen}{The darker, the more correlation exists between the labels.}}
\label{fig:heatmapH7}
\end{figure*}

\boldification{We found labels that appear together (co-occurrence)}

We also found co-occurrence among labels. For instance, ``Test'', ``Logging'', and ``i18n'' appeared often together 
(Figure~\ref{fig:heatmapH7}). This is due to the strong relationship found in the source files. By searching the references for these API-domain categories in the source code, we found ``Test'' in 4,579 source code files, compared to ``Logging'' in 903. The label ``i18n'' appeared in only 73 files. 
On the other hand, the API-domain labels for ``CG'' and ``Security'' usually do not co-occur. ``CG'' only appeared in five java files, while ``Security'' appeared in only 47 files. Future research can investigate co-occurrence techniques to predict co-changes in software artifacts (e.g.,~\cite{wiese2017using}) in this context.

\boldification{In the future, the frequency of co-occurrence could be used to suggest labels}

Figure~\ref{fig:heatmapH7} exhibits the labels' co-occurrence for the dataset containing all the projects. A co-occurrence matrix presents the number of times each label appears in the same context as each possible other label. Examining the aforementioned co-occurrence data, we can determine some expectations and induce some predictions. For example, the ``DB'' label (Database) occurred with more frequency alongside ``Network'' and ``Thread.'' So, it is possible to guess when an issue has both labels, and we likely can suggest a ``Database'' label, even when the machine learning algorithm could not predict it. 
A possible future work can combine the machine learning algorithm proposed in this work with frequent itemset mining techniques, such as apriori~\cite{agrawal1993mining}.

\boldification{The frequency of labels impacts the accuracy}

\textbf{What are the difficulties in labeling accurately?} 
We suspect that the high occurrence of ``UI'', ``Util'', and ``Logic'' labels ($>500$ issues) compared with the low occurrence of ``i18n'', ``Interpreter'', ``GIS'', and ``NLP'' ($<57$ issues) may influence the precision and F-measure values. We tested the classifier with only the top 5 most prevalent API-domain labels and observed no statistically significant differences. One possible explanation is that the transformation method used to create the classifier was Binary Relevance, which creates a single classifier for each label and overlooks possible co-occurrence.

\boldification{The label frequency unbalance occurs due to specific characteristics of the projects}

\responsetoreviewer{RevOneCommentTwoPartC}{The dataset is unbalanced due to the characteristics of the projects. Since JabRef, for instance, is a desktop application, the API-domain label ``UI'' appears more frequently. Table~\ref{tab:confusionM} shows the confusion matrix for the dataset containing all projects (for individual projects, see the appendix). This impacts the prediction of the minor labels even with the SMOTE algorithm, which improves the occurrences of rare labels. Some labels only appear in a few projects. Therefore, even when they are common in a specific project when training and testing with all projects, they may become rare. The recommendation of labels with poor results should be avoided because of the risk of indicating a wrong skill to the contributor.}

\begin{table}[htbp]
 \begin{center}
 \caption{Confusion matrix data and performance from the selected model with all projects}
 \label{tab:confusionM} 
 \begin{tabular}{c|r|r|r|r|r|r} 
 \hline
 \textbf{API-domain} & \textbf{TN} & \textbf{FP} & \textbf{FN} &
 \textbf{TP} & 
 \textbf{Precision} &
 \textbf{Recall} \\
 \hline


 	APM &	125 &	4 &	44 	& 8 & 0.66 & 0.15 \\
 	App &	80 &	22 &	19 &	60 & 0.73 & 0.75\\

 	Big Data &	152 &	0 &	29 &	0 & 0 & 0\\

 	Data Structure &	78 &	24 &	6 &	73 & 0.75 & 0.92\\
 	DB &	163 &	2 &	11 &	5 & 0.71 & 0.31 \\
 	DevOps &	113 &	26 &	0 &	42 & 0.61 & 1\\
 	Error Handling &	97 &	32 &	5 &	47 & 0.59 & 0.90 \\
\hline
 \end{tabular}
 \end{center}
\end{table}

\boldification{While the low frequency affects the prediction, we were able to predict labels with more than 200 occurrences}

Despite the lack of accuracy in predicting the rare labels, we were able to predict those with more than 200 occurrences (all projects together) with reasonable precision (0.84) and/or recall (0.78). We argue the project's nature contributes to the number of issues related to their domain. For example, since the Audacity project is an audio editor and recorder, a high occurrence of ``UI'', ``IO'', and ``Multimedia'' labels is expected. We argue that Audacity's nature contributes to the number of issues related to the labels above. Labels with few samples suffered from low or unstable metrics. ``DB'', for example, varied from 0.09 to 0.9 in recall on predictions depending on the text/train split.

\textbf{Improving the performance of BERT.}
In addition to the number of occurrences of a label, the BERT metrics can be improved by increasing the training set size. Wang et al. \citep{WANG2021107476} and their exploration of several trained deep learning models for GitHub labeling provide important insights into potential performance increases with BERT. The authors showed that the BERT model performed better than the other language models for large datasets with at least 5,000 issues, achieving the highest accuracy, precision, recall, and F-measure scores. However, for small datasets with less than 5,000 issues, CNN outperformed BERT as the best model overall. 
This suggests that BERT depends on the size of the training set of corpus data. 
Therefore, the performance of BERT when labeling GitHub issues will improve with an increased dataset size for the targeted open-source project. \responsetoreviewer{RevThreeCommentTwelve}{When the project datasets were merged (Table \ref{fig:baselineH7}), the BERT metrics decreased the difference from about 26\% to 6\% in precision compared to the other classifiers.}  


\boldification{Experts can also help increase the classification metrics for all models.}

\textbf{What is the impact of the expert classification?} 
Experts can also help increase the classification metrics for all models.
We could observe the C++ project achieved the best F-measure compared with the Java and C\# projects (0.84, 0.82, and 0.80, respectively, with small to large effect sizes). 
Although we evaluated only one C++ project, the results  might suggest after examining Table~\ref{tab:APIclassification} that the number of APIs evaluated by the experts impacts the metrics we will obtain. On the other hand, manual evaluation of a high number of APIs may make generalization unfeasible. The classification carried out by the experts in the C++ project comprised a higher percentage of APIs analyzed. This might be caused by the language characteristics: the libraries' names parsed from the C++ source code had limited information about their use. Thus, classification was more time-consuming. Indeed, the C++ project demanded more effort from the experts to classify it. Ultimately, it became a more detailed classification with better prediction metrics. 

\boldification{Some projects may require less effort from experts than others due to the use of popular APIs}

While experts' analyses are time-consuming, some outlier projects require much less effort than others. For instance, experts analyzed fewer than 3\% of the APIs in RTTS. Since this project imports popular libraries, reuses many libraries across the entire source code and is modular, the expert's work was easier. A possible relationship between popular APIs, modularization, and expert evaluation should be explored in future work. Another possible future work should identify what programming language characteristics impact the expert classification.

\boldification{Reusing APIs can decrease the required effort from experts}

\textbf{To what extent does the proposed method generalize?} 
The semi-automatic classification process decreased the effort carried out by the experts to define the expertise of the APIs. Despite there being considerable effort remaining, as the dataset increases, the rate of new APIs to classify should decrease since projects reuse an average of 35-53\% of core APIs. Third-party libraries account for 8-32\% and 45\% on average (Core + third-party). The use of popular open-source APIs could lead to an impressive 85\% of shared APIs between projects~\cite{qiu2016understanding}.
Farther, the project sizes grow much more quickly than the size of uniquely-used API entities~\cite{qiu2016understanding}. 

\boldification{The demand for expert evaluation should decrease over time as the classifications of one project can transfer to another}

Thus, the demand for expert evaluation should decrease significantly when the number of mined libraries reaches a critical mass (for each programming language), and even new projects may use previous expert evaluations.  
This might impact the method when applied to industry projects, which may use a variety of unique non-free APIs. However, API sharing may happen inside companies or business units, repeating the phenomenon of the libraries' critical mass. Nevertheless, we did not observe this effect, and we could predict labels for an OSS project using data from an industry project. 

\boldification{Another way of reusing previous classifications is through transfer learning}

\textbf{To what extent does the model perform transfer learning?} 
Transfer learning is crucial when projects lack data for training (cold start) or the time or infrastructure to develop their own models. 
This can be particularly problematic in the industry since the data can have restricted access due to security precautions or to comply with procedures or laws. In this situation, the ability to use a pre-trained model is necessary. Using pre-trained models to predict from new data is also desirable because it is faster and cheaper than retraining a model every time a new source project is added to the dataset~\cite{nam2013transfer,seah2013transfer}.
The projects may also benefit from the complementary data from another project when the project dataset is too small for training a predictive model. 

\boldification{However, in our experiments, the transfer learning found a decrease in precision and recall}

The transfer learning experiments found a decrease in precision and recall. The metrics definition: $Precision = \frac{TP}{TP+FP}$ and
$Recall = \frac{TP}{TP+FN}$ indicates the number of False Negatives (FN) and False Positives (FP) that should impact the results. For example, in training and testing individual projects, the RTTS project had a small number of PFs and FNs compared to the transfer learning experiment when RTTS was a target project (Tables \ref{tab:confusionMH1-RMCA} and \ref{tab:confusionMH10-RMCA}).  When targeting the RTTS project, the high number of FNs significantly decreased the recall metric. On the other hand, targeting PowerToys, the number of FPs negatively impacted the precision (Tables~\ref{tab:confusionMH10-RMCA}, \ref{tab:confusionM-powertoys}, and \ref{tab:transfer}). The projects only shared a small number of labels (5 in 31) and are imbalanced among the datasets. For example, ``Setup'' is popular in the RTTS project and rare in JabRef, suggesting the conditional probability distribution of the sources and targets differ. These characteristics might determine which projects match and, therefore, be used to decide the transfer learning source or target. Future work should investigate whether the domain, platform (Web, Desktop, Mobile), architecture, or other project property derives a good match. Furthermore, investigating proxy techniques, such as the one proposed by \citet{nam2013transfer}, to minimize the data distribution difference between target and source projects to predict software engineering defects can be applied to predictions of domain labels of API. \responsetoreviewer{RevOneCommentTwoPartA}{Results for the JabRef (Table \ref{tab:confusionMH10-JABREF}) and Audacity (Table \ref{tab:confusionMH10-AUDACITY}) projects using transfer learning are available in Appendix \ref{sec:AppendixA}. We can observe the high number of FP and FN comparing the Audacity transfer learning results in Table \ref{tab:confusionMH10-AUDACITY} and the results of training and testing the Audacity dataset alone (Table \ref{tab:confusionM-audacity}). Similarly, we can observe the same pattern in the JabRef results in Tables \ref{tab:confusionM-jabref} and \ref{tab:confusionMH10-JABREF}}.

\begin{table}[htbp]
 \begin{center}
 \caption{Confusion matrix and performance. Project RTTS trained/tested alone}
 \label{tab:confusionMH1-RMCA}
 \begin{tabular}{c|r|r|r|r|r|r|r} 
 \hline
 \textbf{API-domain} & \textbf{TN} & \textbf{FP} & \textbf{FN} &
 \textbf{TP} & 
 \textbf{Precision} &
 \textbf{Recall} &
 \textbf{F-measure}\\
 \hline

APM            & 112 & 4  & 2  & 24 & 0.85 &
0.92 &
0.88 \\
Big Data       & 134 & 1  & 2  & 5  & 0.83 &
0.71 &
0.76 \\
Data Structure & 25  & 36 & 3  & 78  & 0.68 &
0.96 &
0.80\\
DB             & 84  & 4  & 19 & 35  & 0.89 &
0.64 &
0.75\\
DevOps         & 60  & 17 & 13 & 52  & 0.75 &
0.80 &
0.77\\
Error Handling & 126 & 5  & 4  & 7   & 0.58 &
0.63 &
0.60 \\
Event Handling & 129 & 0  & 2  & 11  & 1 &
0.84 &
0.91 \\
i18n           & 121 & 6  & 7  & 8  & 0.57 &
0.53 &
0.55 \\
Lang           & 28  & 29 & 11 & 74  & 0.71 &
0.87 &
0.78\\
Logging        & 47  & 24 & 18 & 53  & 0.68 &
0.74 &
0.71\\
Microservices  & 1   & 12 & 0  & 129 & 0.91 &
1 &
0.95\\
Network        & 65  & 20 & 21 & 36  & 0.64 &
0.63 &
0.63\\
Parser         & 69  & 17 & 14 & 42 & 0.71 &
0.75 &
0.73 \\
Security       & 129 & 0  & 8  & 5   & 1 &
0.38 &
0.55\\
Setup          & 66  & 10 & 33 & 33 & 0.76 &
0.50 &
0.60 \\
UI             & 4   & 15 & 0  & 123 & 0.89 &
1 &
0.94\\

\hline
\textbf{Total} & 1200 &	200	& 157& 715 &\multicolumn{3}{l}{}  \\
\hline
 \end{tabular}
 \end{center}
\end{table}

\begin{table}[ht!]
 \begin{center}
 \caption{Confusion matrix and performance: Project RTTS - transfer learning. }
 \label{tab:confusionMH10-RMCA}
 \begin{tabular}{c|r|r|r|r|r|r|r} 
 \hline
 \textbf{API-domain} & \textbf{TN} & \textbf{FP} & \textbf{FN} &
 \textbf{TP} & 
 \textbf{Precision} &
 \textbf{Recall} &
 \textbf{F-measure} \\
 \hline

APM	& 197	& 0	& 38	& 0 & 0 & 0 & 0\\
Data Structure	& 100	& 0	& 135	& 0 & 0 & 0 & 0\\ 
DB	& 145	& 0	& 90	& 0 & 0 & 0 & 0\\
Error Handling	& 223	& 0	& 12	& 0 & 0 & 0 & 0\\
Event Handling	& 212	& 0	& 23	& 0 & 0 & 0 & 0\\
Lang	& 114	& 0	& 121	& 0 & 0 & 0 & 0\\
IO	& 44	& 21	& 131	& 39 & 0.65 & 0.22 & 0.33\\
i18n	& 214	& 0	& 21 & 	0 & 0 & 0 & 0\\
Logging	& 102	& 22	& 91	& 20 & 0.47 & 0.18 & 0.26\\
Logic	& 13	& 17 & 	146 & 	59 & 0.77 & 0.28 & 0.41\\
Microservices	& 28	& 0	& 206	& 1 & 1 & 0.004 & 0.009\\
Network	& 129	& 0	& 106	& 0 & 0 & 0 & 0 \\
Parser	& 156	& 0	& 79	& 0 & 0 & 0 & 0\\
Setup	& 115	& 14	& 98	& 8 & 0.36 & 0.07 & 0.12 \\
Thread	& 175	& 0	& 60	& 0 & 0 & 0 & 0 \\
UI	& 3	& 38	& 26	& 168 & 0.81 & 0.86 & 0.84\\

\hline
\textbf{Total} & 1970 &	112	& 1383 &	295 &\multicolumn{3}{l}{}  \\
\hline

 \end{tabular}
 \end{center}
\end{table}

\begin{table}[ht!]
 \begin{center}
 \caption{Confusion matrix and performance: Project PowerToys - transfer learning.}
 \label{tab:confusionMH10-POWERTOYSOLD}
 \begin{tabular}{c|r|r|r|r|r|r|r} 
 \hline
 \textbf{API-domain} & \textbf{TN} & \textbf{FP} & \textbf{FN} &
 \textbf{TP}  & 
 \textbf{Precision} &
 \textbf{Recall} &
 \textbf{F-measure} \\
 \hline

APM	& 264	& 1	& 88	& 0 & 0 & 0 & 0 \\
App	& 315	& 9	& 28	& 1 & 0.66 & 0.76 & 0.71  \\
Data Structure	& 344	& 3	& 6	& 0 & 0.03 & 0.40 & 0.06 \\
i18n	& 209	& 134	& 4	& 6 & 0 & 0 & 0 \\
Interpreter	& 342	& 1	& 10	& 0 & 0 & 0 & 0 \\
Logging	& 153	& 196	& 0	& 4 & 0.007 & 0.25 & 0.01 \\
Logic	& 173	& 27	& 100	& 53 & 0.01 & 1 & 0.02\\
Microservices	& 0	& 348	& 0	& 5 & 0.25 & 0.49 & 0.33 \\
Parser	& 6	& 105	& 10	& 232 & 0.07 & 0.13 & 0.09 \\
Setup	& 351	& 0	& 2	& 0 & 0.50 & 0.07 & 0.13\\
Test	& 174	& 82	& 56	& 41 & 0.006 & 0.66 & 0.01\\
UI	& 105	& 245	& 1	& 2 & 0.68 & 0.92 & 0.78 \\

\hline
 \end{tabular}
 \end{center}
\end{table}

\boldification{Contributors rate the generated labels differently, according to each project and overall they are positive}

\textbf{How did the contributors rate the labels generated for the issues they solved?}


Overall, participants evaluated the generated labels with positive feedback. The labels classified as important or very important across all the projects were: DevOps (10), DB (5), Services (4), UI (14), Lang (5), Security (1), and Logging (3). Moreover, the labels that were classified as not important were: APM (8), Setup (7), Data Structure (6) and UI (9), and Security (2). 

\boldification{In RTTS, the labels with the best precision were also the best rated ones}

In the RTTS project, all four best-evaluated labels (Services, Logging, Setup, and Network) had precision above 0.75, and two of the four worst-evaluated ones (Data Structure, UI, DevOps, and Lang) had precision $\leq 0.7$. A threshold could determine whether a label must be reported. 

\boldification{However, some participants reported missing labels}

Participants from Cronos projects mentioned they would like to see the label ``Data Structure'' for the evaluated issues. This occurred because we removed the label Data Structure once it was generated for 90\% of the issues selected in the Cronos project. One possibility for that case would be to include in the description of the project that it is strongly based on data structures and that the reported issues likely would involve this knowledge. 

\boldification{Additionally, the participants indicated that they would like to see a clue for the bug's root cause in the labels}

In addition, participants reported some labels could provide a clue for looking for the bug's root cause or determining the work needed to address a new feature request. For Example: \textit{``...some of the labels could give an initial point of view to the reported issue''} (P2) or \textit{``Network: While network tag wasn't that necessary for this particular case, the issue could have been caused by a communication error between the services in which case they would have been''} (P4). On the other hand, some participants preferred not to see more general labels, like Data Structure or Logging, since they are present in many issues:  \textit{``Data Structure is literally everywhere, there wouldn't be any program without them''} (P1), while others missed the Data Structure label (not present in the predicted list because it reached the 90\% threshold) and suggested including it (P14, P16, and P17). Future work can determine how to address developer preferences regarding the inclusion of general labels.

\boldification{The method generalization proposed in this paper also brought problems with labels that do not match completely the projects context}

The generalization of the method proposed in this paper assisted us in embracing more projects. Nevertheless, it also brought problems. We proposed generic labels able to fit a wide range of project types. This might explain the comments about the generic labels. \textit{``...It was a backward compatibility issue with user-defined configuration data, so with a generous interpretation Setup was accurate, but I would have preferred Information Model or Domain Model had it existed''} (P1).
%
%
%
Analyzing the participant's suggestion for a ``Validation'' label, we recollect to the point where the NLP similarity suggested possible API domains for the library related to the issue and the experts' choice. We found the selected API domain was ``Logic'' since no ``Validation'' API-domain label was available. If the experts came from the project, perhaps the API-domain label ``Validation'' could be present and thus meet the participant's needs.

\boldification{The project in Portuguese in this work could also benefit from this generalization since the declared libraries are written in English}

\boldification{Future work can explore the customizability of API-domain labels per project}

Future work can explore more API-domain labels to expand and propose more options to fit additional projects. Customizing labels for the project may generate more precise directions about the skills needed but will require more expert work time. On the other hand, generalization expands the method to a huge range of projects and can decrease the meaning level of the API domains. 


\boldification{Labels can help newcomers review the necessary skills to work on issues}

\textbf{What are the practical implications for different stakeholders?} 
 \MyPara{New contributors.} API-domain labels can help open-source contributors, enabling them to review the skills needed to work on the issues upfront. This is especially useful for new contributors and casual contributors~\cite{Pinto:SANER:2016,AnitaCSCW}, who have no previous experience with the project terminology.

\boldification{Automatic labeling can help maintainers distribute team effort}

\MyPara{Project maintainers.} Automatic API-domain labeling can help maintainers distribute team effort to address project tasks based on required expertise. Project maintainers can also identify which type of APIs generate more issues. Our results show that we can predict the most prominent API domains---in this case ``Util'' and ``Logic''--- with precision up to 95\% and 91\%, respectively (see Table~\ref{tab:confusionM}).

\boldification{Platforms can use our results to redesign the interface and prioritize the most relevant parts for choosing issues}

\MyPara{Platform/Forge Managers.} Participants often selected \textsc{title}, \textsc{body}, and \textsc{labels} to look for information when choosing an issue to which to contribute. 
Our results can be used to propose better layouts for the issue list and detail pages, prioritizing them against other information regions (\ref{fig:hotspotsurvey}). In the issue detail page on GitHub, for instance, the label information appears outside of the main contributor focus, on the right side of the screen. 

\boldification{Templates based on our results can help automated classifiers to use information to predict API labels}

Templates to guide GitHub users in filling out the issues' body to create patterns can be useful in not only making the information space consistent across issues, but also helping automated classifiers that use the information to predict API labels. For instance, some of the wrong predictions in our study could be caused by titles and bodies with little useful information from which to generate labels. 


\boldification{Researchers can extend the proposed approach to other languages and projects}

\MyPara{Researchers.} The scientific community can extend the proposed approach to other languages and projects, including those with more data and different algorithms. Our approach can also be used to improve tools that recommend tasks matched to new contributors' skills and career goals (e.g.,~\cite{sarma2016training}).

\boldification{Educators can assign contributions to OSS as part of their coursework}

\MyPara{Educators.} Educators who assign contributions to OSS as part of their coursework~\cite{pinto2017training} can also benefit from our approach. Labeling issues in OSS projects can help them select examples or tasks for their classes, bringing a practical perspective to the learning environment.

\section{Threats to Validity}

The threats to validity are divided into ``internal,'' ``construct,'' and ``external.'' 

\textbf{Internal Validity.}
One of the threats to the validity of this study is the API domain categorization. We acknowledge the threat that different individuals can create different categorizations, which may introduce some bias in our results. To mitigate this problem, three individuals, including two senior developers and a contributor to the JabRef project, created the API-domain labels categories aiming to generalize to any type of project. 
In the future, we can improve this classification process with a collaborative approach (e.g.,~\cite{8603296,lu2017learning}). 

Although participants with different profiles participated in the JabRef user study, the sample cannot represent the entire population, and the results can be biased. The study randomly assigned a group to each participant. However, some participants did not finish the questionnaire, and the groups ended up lacking balance. Also, the way we created subgroups can introduce bias in the analysis. The practitioners' classification as industry and students were done based on the location of the recruitment, and some students could also be industry practitioners and vice-versa. However, the results of this analysis were corroborated by aggregation according to experience level. 


\textbf{Construct Validity.}
Another concern is the number of issues in our dataset and the link between issues and pull requests. To include an issue/key/tracking ID in the dataset, we linked it to its solution submitted via pull request (or ``revision'' and ``trouble id''). By linking them, we could identify the APIs used to create the labels and define our ground truth (check Section~\ref{sec:DataCollection}). \responsetoreviewer{RevTwoCommentFifteen}{This study does not identify issues merged without PR information.} We manually inspected a random sample of issues (or ``keys'' and ``tracking ids'') to check whether the data was correctly collected and reflected what was shown on the ITS interface. \responsetoreviewer{RevThreeMinorThirteen}{Two authors manually examined 100 tasks randomly picked up from the projects, comparing the collected data with the GitHub interface.} All records were consistent, and all of the issues in this validation set were correctly linked to their pull requests. When the linked data had more than one correspondence, we concatenated all data using the appropriated corpus entry (title, body, comments, description, and summary). Some of the linked data occasionally had repeated text, and can overfit our model. Future versions may improve the data cleaning step. Unlike the other projects, Cronos had multiple linked data through the following columns: ``pai'' and ``ramo,'' ``linked issue'' and ``key,'' and ``key'' and ``ramo.'' This creates a recursive situation where we may link each update with many ``keys'' in different ways. We preferred to keep it simple, using only the linked data that was similar to the other projects: ``key'' and ``ramo.'' 

In prediction models, overfitting occurs when a prediction model exhibits random error or noise instead of an underlying relationship. During the model training phase, the algorithm used information not included in the test set. To mitigate this problem, we also used a shuffle method to randomize the training and test samples. 

Further, we acknowledge that we did not investigate whether the labels helped the users find the most appropriate tasks. It was not part of the user study to evaluate how effective the API labels were in finding a match with user skills. Our focus was on understanding the relevance that the API-domain labels have on the participants' decisions. Besides, we did not evaluate how false positive labels would impact task selection or ranking. However, we believe the impact is minimal since in the three most selected issues, out of 11 recommendations in the JabRef project, only one label was a false positive. \responsetoreviewer{RevThreeCommentTwo}{In addition, when we asked the participants to pick issues with the API labels + project labels (treatment group) or project labels (control group), we might introduce some bias. Indeed, evaluating the difference of relevance perception introduced by the appearance of the new (API-domain) labels should have some influence brought by the poor performance of the project's labels, masking the difference in the measurement experiment. Investigating the effectiveness of API labels by an experiment matching contributors and tasks skills and identifying the problems caused by misclassification are potential avenues for future work.} \responsetoreviewer{RevThreeCommentThree}{The empirical experiment to pick an issue and ask the relevant regions for that choice may introduce a bias since the participant only selected an issue and did not solve the issue.}

When classifying the issues and linked pull requests, we compared the files changed with the parsed source code files at the last version of the projects. If the updated source file is not present anymore, the pull request is discarded.

\textbf{External Validity.}
Generalization is also a limitation of this study. The outcomes could differ for other projects, programming languages ecosystems, or even issues written in a different language. To address this limitation, we extended the previous study \cite{santos2021can} in that direction, mining different projects, including three programming languages, and two \responsetoreviewer{RevThreeMinorTwelve}{natural languages} (or vocabularies). Nevertheless, this study showed how a multi-label classification approach could be useful for predicting API-domain labels and how relevant such a label can be to new contributors. Moreover, the API-domain labels that we identified can generalize to other projects that use the same APIs across multiple project domains (Desktop and Web applications). Many projects adopt a typical architecture (MVC) and frameworks (JavaFX, JUnit, etc.), which makes them similar to many other projects. As described by \citet{qiu2016understanding}, projects adopt common APIs, accounting for up to 53\% of the APIs used. Moreover, our data can be used as a training set for automated API-domain label generation in other projects. 


\section{Conclusion}

We investigated whether API-domain labels are used by  newcomers to select an issue and what information newcomers use to decide what issue to contribute. We found that industry practitioners and experienced coders prefer API-domain labels more often than students and novice coders. Participants prefer API-domain labels over component labels already used in the project. Users would like to see labels with information about issue type, priority, programming language, complexity, technology, and API and pick an issue based on title, body, comments, and labels.


We also investigate to what extent we can predict API-domain labels. We mined data from 22,231 issues from five projects and predicted 31 API-domain labels. Training and testing the projects separately, TF-IDF with the Random Forest algorithm (RF), and unigrams obtained a precision of 84\% and overcame BERT (precision of 62\%). Data from the issue body offered the best results. However, when predicting the API-domain labels for all projects together, RF precision decreased to 78\%, and BERT increased to 72\%, suggesting the positive sensibility of the BERT technique when applied to larger datasets. 

Transferring learning from diverse sources and targets resulted in a decrease in evaluation metrics with an extensive range of values regarding precision and recall. Future work should investigate ways to determine when or how to apply transfer learning to API-domain labels among projects.

Finally, developers agreed that up to 64.4\% of the API-domain labels are important to identify the skills and therefore should help to solve the issues if they are available.


This study is a step toward helping new contributors match their API skills with each task and better identify an appropriate task to start their onboarding process into an OSS project. 

\section*{Acknowledgment}

This work is partially supported by the National Science Foundation under Grant numbers 1815486, 1815503, 1900903, and 1901031, CNPq grant \#313067/2020-1. We also thank the developers who spent their time answering our questionnaire.

\section*{Data availability}

The datasets generated during and/or analyzed during the current study are available in the zenodo repository\footnote{\responsetoreviewer{RevThreeCommentOne}{\url{https://doi.org/10.5281/zenodo.6869246}}}.

\section*{Declarations}

\textbf{Conflicts of interest/competing interests} The authors declare that they have no conflict of interest.

\bibliographystyle{IEEEtranN}
\bibliography{msr}

\appendix
\newpage

\section{Appendix}
\label{sec:AppendixA}

\textbf{Additional data from RQ2 results.}
Some data were presented with box plots in section \ref{sec:RQ2-results}. The redundant data (and more detailed) about the experiments are available here in tables.

\begin{table}[ht!]
 \centering
 \caption{Overall performance from models created to evaluate the corpus. Hla* }
 \label{tab:resultsTFIDF-BERT}
 \begin{tabular}{c|c|r|r|r|r} 
  \hline

  \textbf{Model} & \textbf{Corpus} & \textbf{Precision} & \textbf{Recall} & \textbf{F-measure} &
  \textbf{Hla} \\
  \hline
TD-IDF & Title (T) 
& 0.830 & 	0.794 & 	0.809 & 0.117 \\
TD-IDF & Body (B)
& 0.840 & 	0.786 & 	0.811 & 0.116\\
TD-IDF & T, B 
& 0.839 & 	0.799 & 	0.817 & 0.113\\
TD-IDF & T, B, Comments 
& 0.831 &	0.796 &	0.812 & 0.116\\
BERT & Title (T) 
& 0.616 & 	0.592 & 	0.596 & 0.277 \\
BERT & Body (B)
& 0.599 & 	0.598 & 	0.591 & 0.27\\
BERT & T, B 
& 0.595 & 	0.559 & 	0.568 & 0.269\\
BERT & T, B, Comments 
& 0.597 &	0.587 &	0.582 & 0.266\\
 \hline
\multicolumn{5}{l}{\scriptsize\textit{* Hla - Hamming Loss }}
 \end{tabular}
\end{table}

\begin{table}[ht!]
 \begin{center}
 \caption{Overall performance from models created to evaluate the number of grams. Hla*  }
 \label{tab:resultsGrams}
 \begin{tabular}{c|r|r|r|r} 
  \hline

  \textbf{Model} & \textbf{Precision} & \textbf{Recall} & \textbf{F-measure} &
  \textbf{Hla*} \\
  \hline
unigrams (1,1)

& 0.841 & 	0.829 & 	0.834 & 0.115 \\
bigrams (2,2)

& 0.844 & 	0.809 & 	0.825 & 0.119\\
trigrams (3,3)

& 0.841 & 	0.809 & 	0.822 & 0.123\\
quadrigrams (4,4)

& 0.845 &	0.798 &	0.819 & 0.125\\
 \hline
\multicolumn{5}{l}{\scriptsize\textit{* Hla - Hamming Loss }}
\\
 \end{tabular}
 \end{center}
\end{table}

\begin{table}[ht!]
 \begin{center}
 \caption{Overall performance from models created to evaluate the algorithms. Hla* }
 \label{tab:resultsAlg}
 \begin{tabular}{c|r|r|r|r} 
  \hline

  \textbf{Model} & 
  \textbf{Hla} & \textbf{Precision} & \textbf{Recall} & \textbf{F-measure}  \\
  \hline
DecisionTree &	 	0.105 & 	0.861 &	0.837 &	0.847 	\\
Dummy 	&	0.202 &	0.749 &	0.658 &	0.698 \\
LogisticRegression 	&	0.120 	&	0.858 &	0.792 &	0.822 \\	
MLPClassifier 	&	0.107 &	0.853 &	0.846 &	0.848 \\
MLkNN &		0.126 &	0.837 &	0.801 &	0.816 \\	
RandomForest 	& 	0.107 &	0.864 &	0.836 &	0.849 \\
BERT                &   0.277   &  0.601 & 0.574 & 0.578 \\
 \hline
\multicolumn{5}{l}{\scriptsize\textit{* Hla - Hamming Loss }}
 \end{tabular}
 \end{center}
\end{table}

\begin{table}[ht!]
 \begin{center}
 \caption{Overall performance from models created using the dataset with all projects merged to evaluate the algorithms}
 \label{tab:resultsAllTogether}
 \begin{tabular}{c|r|r|r|r} 
  \hline

  \textbf{Model} &
  \textbf{Hla*} & \textbf{Precision} & \textbf{Recall} & \textbf{F-measure}  \\
  \hline
DecisionTree        &	0.157 	&	0.768 &	0.576 &	0.654 	\\
LogisticRegression  &	0.167   &	0.779 &	0.504 &	0.611 	\\
MLPClassifier       &	0.154 	&	0.766 &	0.595 &	0.666 \\
MLkNN 	 	        &   0.179   &	0.709 &	0.555 &	0.617 \\
RandomForest        &	0.153   &	0.785 &	0.573 &	0.659\\
BERT & 0.219 & 0.725 & 	0.511 & 	0.593 \\
 \hline
\multicolumn{5}{l}{\scriptsize\textit{* Hla - Hamming Loss }}
\\
 \end{tabular}
 \end{center}
\end{table}

\responsetoreviewer{RevOneCommentTwoPartB}{We also include the confusion matrix for all projects trained and tested alone (Tables~\ref{tab:confusionM-jabref}--\ref{tab:confusionMH10-AUDACITY}). The confusion matrix for the RTTS project is in Table \ref{tab:confusionMH1-RMCA} on Section \ref{sec:discussion}.}

\begin{table}[htbp]
 \begin{center}
 \caption{Overall performance from the selected model - JabRef project}
 \label{tab:confusionM-jabref} 
 \begin{tabular}{c|r|r|r|r|r|r} 
 \hline
 \textbf{API-domain} & \textbf{TN} & \textbf{FP} & \textbf{FN} &
 \textbf{TP} & 
 \textbf{Precision} &
 \textbf{Recall} \\
 \hline
 	Network &	13 	& 6 &	8 &	19 & 0.76 & 0.70\\
 	DB 	& 43 	&1 &	0 &	2 & 0.67 & 1\\
	Interpreter &	12& 	9& 	5& 	20 & 0.69 & 0.8\\
 	Logging &	0& 	7& 	0& 	39 & 0.85 & 1\\
 	Data Structure& 	45& 	0& 	0& 	1 & 1 & 1\\
 	i18n &	40& 	0& 	5& 	1 & 1 & 0.17\\
 	Setup &	33& 	3& 	1& 	9 & 0.75 & 0.9\\
 	Microservices &	42& 	0& 	4& 	0 & 0 & 0\\
 	Test &	41& 	0& 	3& 	2 & 1 & 0.4\\
 	IO &	0& 	6& 	0& 	40 & 0.87 & 1\\
	UI &	4& 	2& 	0& 	40 & 0.95 & 1\\
 	App &	41& 	1& 	2& 	2 & 0.67 & 1\\
\hline
 \end{tabular}
 \end{center}
\end{table}

\begin{table}[htbp]
 \begin{center}
 \caption{Overall performance from the selected model - Powertoys project}
 \label{tab:confusionM-powertoys} 
 \begin{tabular}{c|r|r|r|r|r|r} 
 \hline
 \textbf{API-domain} & \textbf{TN} & \textbf{FP} & \textbf{FN} &
 \textbf{TP} & 
 \textbf{Precision} &
 \textbf{Recall} \\
 \hline
        APM &	72  &	3 & 	11 & 	7 & 	0.70 & 	0.39\\
 	Interpreter & 	91 & 	0 & 	1 & 	1 & 	1 & 	0.50\\
 	Logging & 	92 & 	1 & 	0 & 	0 & 	0  &	0\\
 	Data Structure & 	90 & 	1 & 	0 & 	2 & 	0.67 & 	1\\
 	i18n & 	92 & 	0 & 	1 & 	0 & 0 & 	0\\
 	Setup & 	47 & 	6 & 	11 & 	29 & 	0.83 & 	0.72\\
 	Logic & 	87 & 	1 & 	0 & 	5 & 	0.83 & 	1\\
 	Microservices & 	70 & 	1 & 	15 & 	7 & 	0.88  &	0.32\\
 	Test & 	91 & 	1 & 	0 & 	1 & 	0.50 & 	1\\
 	Search  &	41 & 	6 & 	8 & 	38 & 	0.86 & 	0.83\\
        UI & 	25 & 	15 & 	1 & 	52 & 	0.78 & 	0.98\\
	Parser & 	87 & 	1 & 	2 & 	3 & 	0.75 & 	0.60\\
 	App & 	22 & 	12 & 	0 & 	59 & 	0.83 & 	1\\
\hline
 \end{tabular}
 \end{center}
\end{table}

\begin{table}[htbp]
 \begin{center}
 \caption{Overall performance from the selected model - Audacity project}
 \label{tab:confusionM-audacity} 
 \begin{tabular}{c|r|r|r|r|r|r} 
 \hline
 \textbf{API-domain} & \textbf{TN} & \textbf{FP} & \textbf{FN} &
 \textbf{TP} & 
 \textbf{Precision} &
 \textbf{Recall} \\
 \hline
Util           & 38 & 3 & 2 & 13 & 0.81 & 0.87 \\
APM            & 50 & 1 & 2 & 3  & 0.75 & 0.60 \\
Network        & 54 & 0 & 0 & 2  & 1.00 & 1.00 \\
DB             & 49 & 1 & 2 & 4  & 0.80 & 0.67 \\
Error Handling & 37 & 3 & 2 & 14 & 0.82 & 0.88 \\
Logging        & 52 & 0 & 1 & 3  & 1.00 & 0.75 \\
Thread         & 46 & 1 & 0 & 9  & 0.90 & 1.00 \\
Lang           & 54 & 0 & 0 & 2  & 1.00 & 1.00 \\
Data Structure & 15 & 8 & 2 & 31 & 0.79 & 0.94 \\
i18n           & 48 & 2 & 0 & 6  & 0.75 & 1.00 \\
Setup          & 13 & 5 & 2 & 36 & 0.88 & 0.95 \\
Logic          & 3  & 0 & 0 & 53 & 1.00 & 1.00 \\
IO             & 10 & 3 & 6 & 37 & 0.93 & 0.86 \\
UI             & 6  & 2 & 0 & 48 & 0.96 & 1.00 \\
Parser         & 51 & 1 & 0 & 4  & 0.80 & 1.00 \\
Event Handling & 28 & 6 & 1 & 21 & 0.78 & 0.95 \\
App            & 29 & 4 & 4 & 19 & 0.83 & 0.83 \\
GIS            & 50 & 1 & 2 & 3  & 0.75 & 0.60 \\
Multimedia     & 15 & 4 & 5 & 32 & 0.89 & 0.86 \\
CG             & 50 & 0 & 1 & 5  & 1.00 & 0.83 \\
\hline
 \end{tabular}
 \end{center}
\end{table}

\begin{table}[htbp]
 \begin{center}
 \caption{Overall performance from the selected model - MTT project}
 \label{tab:confusionM-cronos} 
 \begin{tabular}{c|r|r|r|r|r|r} 
 \hline
 \textbf{API-domain} & \textbf{TN} & \textbf{FP} & \textbf{FN} &
 \textbf{TP} & 
 \textbf{Precision} &
 \textbf{Recall} \\
 \hline
 NLP      & 34 & 0 & 9 & 9  & 1.00 & 0.50 \\
APM      & 9  & 0 & 0 & 43 & 1.00 & 1.00 \\
DB       & 37 & 0 & 4 & 11 & 1.00 & 0.73 \\
Lang     & 10 & 1 & 2 & 39 & 0.97 & 0.95 \\
DevOps   & 0  & 5 & 0 & 47 & 0.90 & 1.00 \\
Setup    & 18 & 6 & 8 & 20 & 0.77 & 0.71 \\
IO       & 43 & 0 & 4 & 5  & 1.00 & 0.56 \\
UI       & 0  & 2 & 0 & 50 & 0.96 & 1.00 \\
Security & 9  & 4 & 2 & 37 & 0.90 & 0.95 \\
\hline
 \end{tabular}
 \end{center}
\end{table}

\begin{table}[ht!]
 \begin{center}
 \caption{Confusion matrix and performance: Project JabRef - transfer learning. }
 \label{tab:confusionMH10-JABREF}
 \begin{tabular}{c|r|r|r|r|r|r} 
 \hline
 \textbf{API-domain} & \textbf{TN} & \textbf{FP} & \textbf{FN} &
 \textbf{TP} & 
 \textbf{Precision} &
 \textbf{Recall} \\
 \hline

Network        & 65  & 7  & 43 & 3  & 0.30 & 0.07 \\
DB             & 107 & 8  & 3  & 0  & 0    & 0    \\
Interpreter    & 71  & 0  & 47 & 0  & 0    & 0    \\
Logging        & 45  & 3  & 63 & 7  & 0.70 & 0.10 \\
Data Structure & 98  & 16 & 4  & 0  & 0    & 0    \\
i18n           & 103 & 0  & 15 & 0  & 0    & 0    \\
Setup          & 43  & 72 & 0  & 3  & 0.04 & 1    \\
Microservices  & 77  & 24 & 12 & 5  & 0.17 & 0.29 \\
Test           & 77  & 13 & 23 & 5  & 0.28 & 0.18 \\
IO             & 33  & 0  & 67 & 18 & 1.00 & 0.21 \\
UI             & 2   & 35 & 13 & 68 & 0.66 & 0.84 \\
App            & 9   & 90 & 0  & 19 & 0.17 & 1   \\
\hline

 \end{tabular}
 \end{center}
\end{table}

\begin{table}[ht!]
 \begin{center}
 \caption{Confusion matrix and performance: Project Audacity - transfer learning. }
 \label{tab:confusionMH10-AUDACITY}
 \begin{tabular}{c|r|r|r|r|r|r} 
 \hline
 \textbf{API-domain} & \textbf{TN} & \textbf{FP} & \textbf{FN} &
 \textbf{TP} & 
 \textbf{Precision} &
 \textbf{Recall} \\
 \hline
APM            & 127 & 1  & 9   & 0   & 0    & 0    \\
Network        & 132 & 1  & 4   & 0   & 0    & 0    \\
DB             & 117 & 0  & 20  & 0   & 0    & 0    \\
Error Handling & 105 & 0  & 32  & 0   & 0    & 0    \\
Logging        & 88  & 41 & 7   & 1   & 0.02 & 0.13 \\
Thread         & 122 & 0  & 15  & 0   & 0    & 0    \\
Lang           & 133 & 0  & 4   & 0   & 0    & 0    \\
Data Structure & 37  & 0  & 100 & 0   & 0    & 0    \\
i18n           & 118 & 0  & 19  & 0   & 0    & 0    \\
Setup          & 28  & 1  & 94  & 14  & 0.93 & 0.13 \\
Logic          & 5   & 10 & 60  & 62  & 0.86 & 0.51 \\
IO             & 31  & 8  & 65  & 33  & 0.80 & 0.34 \\
UI             & 0   & 18 & 9   & 110 & 0.86 & 0.92 \\
Parser         & 126 & 1  & 10  & 0   & 0    & 0    \\
Event Handling & 68  & 0  & 69  & 0   & 0    & 0    \\
App            & 58  & 33 & 12  & 34  & 0.51 & 0.74 \\
\hline

 \end{tabular}
 \end{center}
\end{table}

\end{document}